\documentclass[9pt,twocolumn,twoside,lineno]{article}
\usepackage[skip=0pt plus0pt, indent=20pt]{parskip}

\usepackage[left=0.5in,right=0.5in,top=0.5in,bottom=0.5in]{geometry}
\AtEndOfClass{\RequirePackage[nopatch=eqnum]{microtype}}
\usepackage{abstract}

\RequirePackage[utf8]{inputenc}
\RequirePackage[english]{babel}
\RequirePackage{amsmath,amsfonts,amssymb}
\RequirePackage{lmodern}
\RequirePackage[scaled]{helvet}
\RequirePackage{lettrine} % For dropped capitals

\usepackage[numbers,sort&compress]{natbib} % optional but recommended
\usepackage{url}                     

\RequirePackage{afterpage}
\RequirePackage{ifpdf,ifxetex}
\ifpdf\else
  \ifxetex\else
    
    \pdfpagewidth=\paperwidth
    \pdfpageheight=\paperheight
\fi\fi
\RequirePackage{xcolor}
\RequirePackage{tikz}
\RequirePackage[framemethod=tikz]{mdframed}

\usepackage{hyperref}
\hypersetup{colorlinks,allcolors=blue}

\RequirePackage{authblk}
\usepackage{comment}
\title{Deformation Driven Suction Cups: A Mechanics-Based Approach to Wearable Electronics}

\author[a,b]{Seola Lee}
\author[a]{Andrew Akerson}
\author[a]{Roham Pardakhtim}
\author[a]{Ehsan Hajiesmaili}
\author[a]{Kevin Rhodes}
\author[a]{Zidong Li}
\author[a]{Andrew Stanley}
\author[a]{Amirhossein Amini}
\author[a]{Daniele Piazza}
\author[a,b]{Chiara Daraio}
\author[a,*]{Tianshu Liu}

\affil[a]{\small Reality Labs Research, Meta Platforms, Inc., Redmond, WA 98052, USA}
\affil[b]{\small Division of Engineering and Applied Science, California Institute of Technology, Pasadena, CA 91125, USA}
\affil[*]{\small Corresponding author: Tianshu Liu, Tianshu.liu@meta.com}

\date{}

\begin{document}

\twocolumn[
  \begin{@twocolumnfalse}
    \maketitle
    \begin{abstract}
    Wearable electronics are emerging as essential tools for health monitoring, haptic feedback, and human-computer interactions. While stable contact at the device–body interface is critical for these applications, it remains challenging due to the skin’s softness, roughness, and mechanical variability. Existing methods, such as grounding structures or adhesive tapes, often suffer from contact loss, limited repeatability, and restrictions on the types of electronics they can support. Suction-based adhesives offer a promising alternative by generating negative pressure without requiring tight bands or chemical adhesives. However, most existing cup designs rely on rigid-surface assumptions and overlook mechanical interactions between suction cups and skin. Inspired by traditional cupping therapies, we present a suction-based adhesive system that attaches through elastic deformation and recovery. Using analytical modeling, numerical simulations, and experiments, we present a mechanics-based framework showing how suction performance depends on cup geometry, substrate compliance, and interfacial adhesion. We show that cup geometry should be tailored to substrate stiffness. Wide, flat suction cups perform well on rigid surfaces but fail on soft ones like skin due to substrate intrusion into the chamber. Narrow and tall domes better preserve recoverable volume and generate stronger suction. To improve sealing on rough, dry skin, we introduce a soft, tacky interfacial layer informed by a contact mechanics model. Using our design principles for skin suction adhesives, we demonstrate secure attachment of rigid and flexible components including motion sensors, haptic actuators, and electrophysiological electrodes across diverse anatomical regions. These findings provide a fundamental basis for designing the next generation of skin-friendly adhesives for wearable electronics.
    \end{abstract}
     \vspace{1em} 
  \end{@twocolumnfalse}
]
Wearable electronics are rapidly evolving as platforms for continuous health monitoring~\cite{brasier_applied_2024, emaminejad_autonomous_2017}, user intent detection~\cite{labs_generic_2024, maereg_hand_2025, zhou_sign--speech_2020,tchantchane_review_2023,grauman_ego4d_2022, grauman_ego-exo4d_2024, david-john_towards_2021}, and immersive human-computer interaction~\cite{pezent_tasbi_2019, venkata_electro-elastic_2023,xiong_augmented_2021,licklider_man-computer_1960}. These systems are expected to operate across diverse regions of the body while remaining functional under motion. However, maintaining reliable contact with soft, irregular, and dynamically varying skin remains a core challenge.

Commercial devices, such as smartwatches and virtual reality (VR) headsets, typically rely on grounding structures of straps or bands (Fig.~\ref{figure1}\textit{A}). While these are effective at localized attachment, they can only be used on specific regions of the body and are prone to frequent contact loss during movement~\cite{bottcher_data_2022}. Medical-grade wearables, including electrocardiogram (ECG) and  electromyogram (EMG) systems, use skin adhesives to enhance stability, but are generally limited to single use and may cause irritation or leave adhesive residues~\cite{ray_bio-integrated_2019}. Emerging platforms such as electronic tattoos (E-tattoos) and skin-conformal electronics offer promising form factors for continuous wear~\cite{chortos_pursuing_2016, li_e-tattoos_2024}, yet they often require specialized fabrication and are incompatible with rigid or macro-scale electronics (Fig.~\ref{figure1}\textit{A}).

Among mechanical adhesion strategies~\cite{boesel_gecko-inspired_2010, baik_highly_2018, lee_softened_2023}, suction-based adhesives have gained attention for their simplicity, scalability, and strong normal attachment~\cite{veggel_classification_2024}. Industrial vacuum cups are widely used to handle smooth, flat surfaces such as silicon wafers and glass panels. More recently, octopus-inspired suction devices have shown strong underwater performance by utilizing compliant structures and fluid-assisted sealing mechanisms~\cite{veggel_classification_2024, yue_bioinspired_2024, baik_wet-tolerant_2017}.

These bio-inspired, suction interfaces have also been explored for skin-interfacing electronics, demonstrating strong potential for power-free, reversible adhesion without the need for active vacuum systems or adhesives~\cite{lee_softened_2023, baik_highly_2018,huang_suction_2021,chun_conductive_2018}. Although these designs have shown promise, it is unclear how the cup geometry, interface conditions, and substrate properties affect adhesion with skin. Most contemporary designs remain qualitatively inspired and lack a rigorous understanding of the mechanics that govern suction performance, particularly the coupled interactions between the suction cup and the substrates. On flat, rigid surfaces, these interactions are minimal. However, on soft, irregular substrates such as skin, the mechanical properties such as compliance, surface roughness, and adhesion play a critical role in suction behavior. Additionally, many existing designs rely on flexible backing layers, which limits their compatibility with rigid electronic components. In the absence of a systematic design framework, performance across different geometries and substrates cannot be easily predicted or generalized.

In this work, we address this knowledge gap by developing a mechanics-based framework for the design of suction-based adhesive on soft, skin-like substrates. Inspired by traditional Chinese cupping therapies that achieve reliable adhesion to the skin through passive elastic deformation, we combine theoretical modeling, numerical simulations, and experiments to reveal how the aspect ratio of the suction cup and the stiffness of the substrate together determine pressure differentials and adhesion strength. To minimize leakage, we introduce a soft, tacky interfacial layer that enhances pressure distribution and improves sealing. We demonstrate that both mechanical compliance and the interfacial work of adhesion are critical for sustaining attachment. This study provides fundamental insights into suction mechanics on soft surfaces and establishes design principles for scalable, re-attachable suction-based adhesives for skin. By enabling stable attachment across a wide range of skin regions including fingertips, palms, nails, and wrists, our approach supports the integration of both flexible and rigid electronics for next-generation health monitoring and human-machine interfaces.

%%%%%%%%%%----------------------Results-----------------------%%%%%%%%%
\section*{Results}
\subsection*{Mechanisms and Designs}
\begin{figure*}
\centering
    \includegraphics[width=0.9\textwidth]{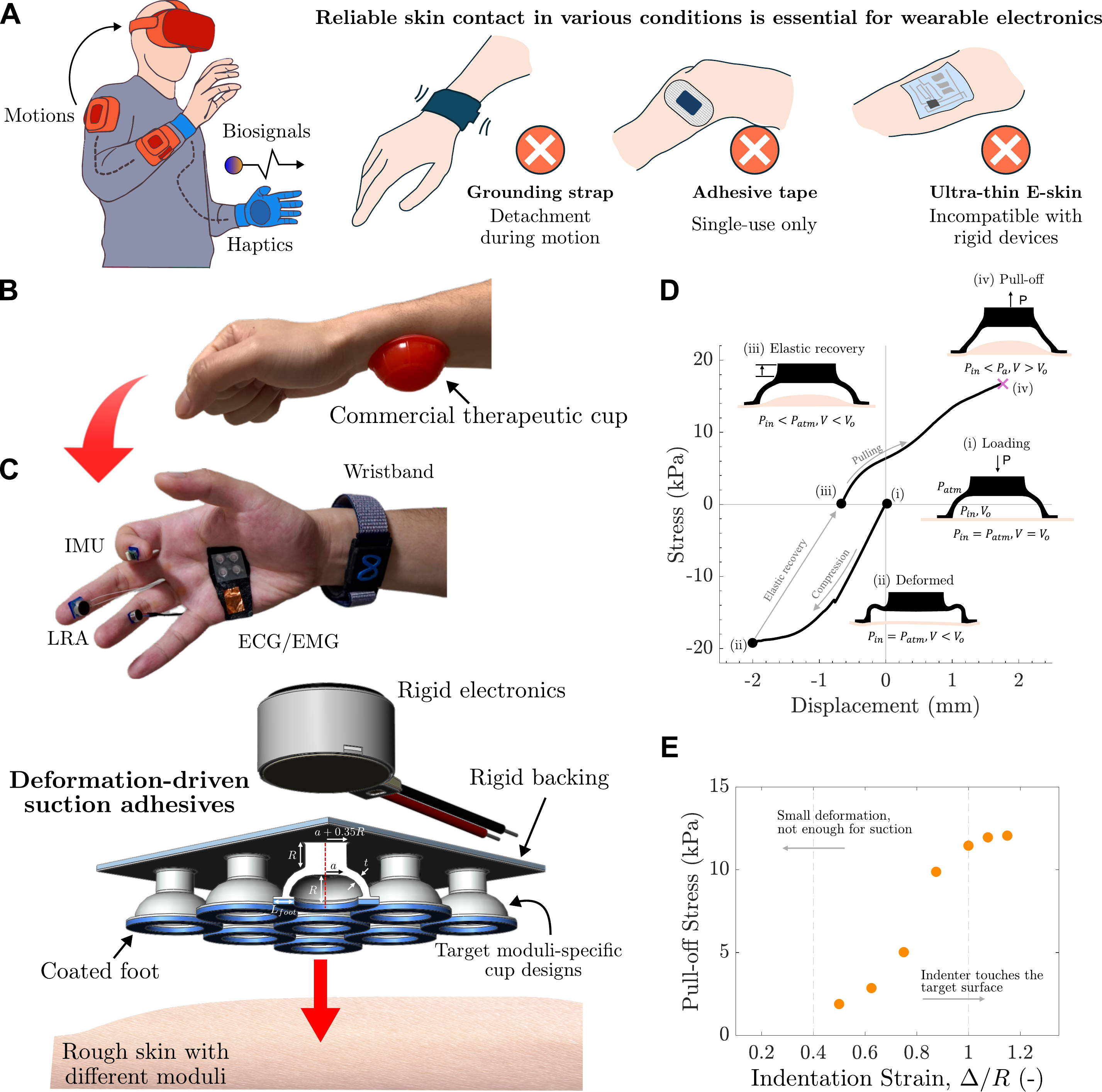}
    \caption{\textbf{Suction Cup Design and Mechanisms.} \textbf{(\textit{A})} Schematic of conventional wearable electronics and traditional adhesive strategies. While reliable skin contact is essential for device performance, existing solutions face trade-offs. \textbf{(\textit{B})} Inspiration from traditional cupping therapy and an example of a deformation-driven suction cup attached to skin. \textbf{(\textit{C})} Proposed deformation-driven suction cup adhesives transforming various electronic devices into wearables. Suction cup design, key geometrical parameters including cup width ($a$) cup thickness ($t$) and cup size ($R$), and coated footing layers are optimized for rough skin with varying stiffness. A stiff backing layer enables integration with rigid electronics. \textbf{(\textit{D})} Suction is activated through 2 steps: compression (i-ii) and elastic recovery (ii-iii). This process expels air and creates pressure difference between inside and outside of the cup, which enables a finite pull-off force (iii-iv). The data reported here is based on $R=2$ mm, $a/R=2$, $t/R=0.3$ cup on a PDMS substrate. \textbf{(\textit{E})} The extent of indentation ($\Delta$) is critical to pull-off stress. Higher initial indentation strain ($\Delta/R$) leads to higher pull-off stress. Data reported here is based on $R=2$ mm, $a/R=1$, $t/R=0.3$.}
    \label{figure1}
    \index{figures}
\end{figure*}

Cupping therapies have been practiced for centuries as a means of attaching domed cups to the skin without the need for adhesives or straps. These traditional devices achieve long-duration adhesion by creating negative pressure within the cup, drawing soft tissues upward and maintaining contact through a vacuum seal. Among various activation mechanisms, such as heat-induced~\cite{lee_octopus-inspired_2016} air expansion or external pumping~\cite{frey_octopus-inspired_2022,yue_bioinspired_2024}, our work focuses on deformation-driven suction. Here, elastic recovery alone generates the vacuum pressure. This strategy requires no external inputs and is widely used in commercial silicone therapy cups (Fig.~\ref{figure1}\textit{B}), which adhere to the body through manual compression and release.

Inspired by this principle, we develop a deformation-driven suction adhesive system that is designed to allow robust attachment of rigid and flexible electronic components across various regions of the body (Fig.~\ref{figure1}\textit{C}). Our approach couples elastic shell deformation with controlled cavity volume change to achieve reliable skin contact. The suction cup geometry is abstracted from therapeutic cupping devices, featuring a hemispherical dome with a flat sealing rim. Key geometric parameters include the cup radius $R$, shell thickness $t$, and the roof width $a$, as shown in Fig.~\ref{figure1}\textit{C}. During activation, the cup undergoes elastic deformation and recovery through shell buckling, allowing the skin to deform upward into the inner cavity to generate suction. A larger $a$ results in a flatter dome with a broader central area, while $a=0$ corresponds to a hemispherical geometry without the flat roof. Thicker shells (larger $t$) offer greater structural resistance and recovery from deformation, while thinner shells are more easily compressed. The design also incorporates a backing layer and a coated soft foot to facilitate integration with wearable devices and ensure conformal contact on rough skin with varied stiffness.

To enable controlled shell buckling during loading, we include an indenter structure to the apex of the dome\footnote{The indenter's properties are not considered in this study; however, it remains a necessary feature to localize stress and initiate elastic deformation.}. The indenter radius is set to $a+0.35R$, ensuring uniform compression over the flat central region. In addition, we introduce a widened footing to enhance contact and peripheral sealing, with thickness $t_{foot} = 0.12R$ and lateral width $L_{foot} = 0.35R$. All geometries are designed to be easily scalable, with the cup radius $R$ used as the characteristic length scale throughout our study.
 
As shown in Fig.~\ref{figure1}\textit{D}, the suction activation process begins with the user applying a compressive load $P$ onto the cup. The deformation expels air from the cavity and reduces the enclosed volume $V$, while the internal pressure $P_{in}$ remains approximately equal to the atmospheric pressure $P_{atm}$ from the cup-skin boundary equilibrium condition. Upon the sudden release of the load (Fig.~\ref{figure1}\textit{D}, ii), the cup rapidly attempts to return to its original shape, increasing the enclosed volume. Under isothermal conditions, this increase leads to a drop in internal pressure following the ideal gas law
\begin{equation}
    P_{in}V = const.,
\end{equation}
forming a vacuum relative to $P_{atm}$. Unlike the compression stage (i), this stage results in inward and downward forces at the cup-skin interface, improving sealing and resisting air leakage. The resulting suction force (iii) resists detachment (iv) until a critical pulling load is reached,
\begin{equation}
    P = -P_{pull-off} \approx -(P_{atm}-P_{in}) \pi (a+R)^2,
\end{equation}
where the cup-skin seal breaks and vacuum effects vanish.

As illustrated in Fig.~\ref{figure1}\textit{E}, increasing the indentation strain in stage (i), defined as the ratio of indenter displacement $\Delta$ to the cup size $R$, leads to a stronger suction force, $P_{pull-off}$. This is due to the larger volume change during the elastic recovery phase, which yields a lower equilibrium internal pressure $P_{in}$. Below a strain of 0.3, no stable suction is observed, and reliable pull-off forces could not be measured. At high strains exceeding 1, where the indenter compresses the cup to its full height, the indenter contacts the substrate, and the $P_{pull-off}$ plateaus. These observations confirm that suction strength is directly related to the volume change achieved during activation. To allow the full deformation of the cup, we set the indenter height equal to the cup size $R$.

To complement the experimental measurements, we develop a modeling framework to simulate the suction process. This allows us to computationally evaluate the key physical quantities governing suction performance, that is, the internal pressure $P_{in}$, the enclosed volume $V$, and the pull-off stress $P_{pull-off}$. We construct a variational model through a total potential energy function $\Pi$ which includes contributions from the work of the ideal gas and the large-deformation elastic energies of both the suction cup and skin, 
\begin{equation}
    \begin{split}
        \Pi &=\int_{\Omega_{cup}} \hspace{-1em} W_{cup}(F) \ d\Omega + \int_{\Omega_{skin}} \hspace{-1em} W_{skin}(F) \ d \Omega \\
        & \quad - P_{atm} V_0 \log{\frac{V}{V_0}} + P_{atm} (V - V_0),
    \end{split}
\end{equation}
where $W_{cup}$ and $W_{skin}$ are the strain energy densities of the cup and skin, $F$ is the deformation gradient tensor, and $V_0$ is the enclosed volume when suction is initiated. Minimizing this energy gives a set of equilibrium relations, which we discretize through the Finite Element Method (FEM). We develop a custom solver built on the open-source deal.ii FEM library~\cite{africadeal2024}, and full details of the modeling and numerics are provided in the~\textit{SI Appendix, Section 3}.

\subsection*{Effects of Skin Compliance}
Surfaces of the human body span a broad range of mechanical stiffness. Fingernails and other stiff regions can have a stiffness of up to $\sim$3 GPa~\cite{tohmyoh_nanoindentation_2023}, while highly compliant areas such as the forearm ($\sim$100 kPa)~\cite{liang_biomechanical_2010}, palms ($\sim$30 kPa)~\cite{liang_biomechanical_2010} and fingertips ($\sim$10s of kPa)~\cite{abdouni_biophysical_2017} are orders of magnitude softer. Even within a single anatomical region, the stiffness of the individual skin layers can vary by over three times across the population~\cite{pailler-mattei_vivo_2008, crichton_elastic_2013, wei_allometric_2017}. These heterogeneous layers include the outer epidermis ($\sim$4 MPa), the middle dermis ($\sim$40 kPa) and the deep hypodermis ($\sim$15 kPa), each with different mechanical and biological functions~\cite{feng_vivo_2022}. The thickness and mechanical properties of each layer varies significantly with age, anatomical location, and hydration. Overall, this makes skin a highly heterogeneous, compliant, viscoelastic substrate~\cite{silver_viscoelastic_2001,khatyr_model_2004,ruvolo_jr_skin_2007}. Therefore, designing suction adhesives for wearable electronics requires accommodating this diversity. Most existing vacuum cup designs assume rigid substrates and neglect deformation of the target surface. While this assumption holds for hard surfaces, it breaks down for skin-like substrates whose effective modulus can be comparable to that of the cup. In such cases, the substrate deformation becomes substantial, altering the suction mechanics and requiring a revised design strategy.

Using both experiments and finite element simulations, we systematically investigate how suction performance varies with substrate stiffness and the cup goemetry. We vary two geometric parameters: the roof width $a$ and the shell thickness $t$ (defined in Fig.~\ref{figure1}\textit{C}). Suction cups are tested on substrates ranging from rigid silicon wafers ($\sim$100 GPa) to highly compliant silicone elastomers including PDMS 10:1 ($\sim$1 MPa), Ecoflex 00-20 ($\sim$40 kPa), and Ecoflex Gel ($\sim$20 kPa), covering stiffness ratios relevant to wearable applications. To decouple substrate stiffness from surface properties, we coat all substrates with a thin layer of PDMS to ensure they have identical surface properties.
 
\begin{figure*}
\centering
    \includegraphics[width=0.9\textwidth]{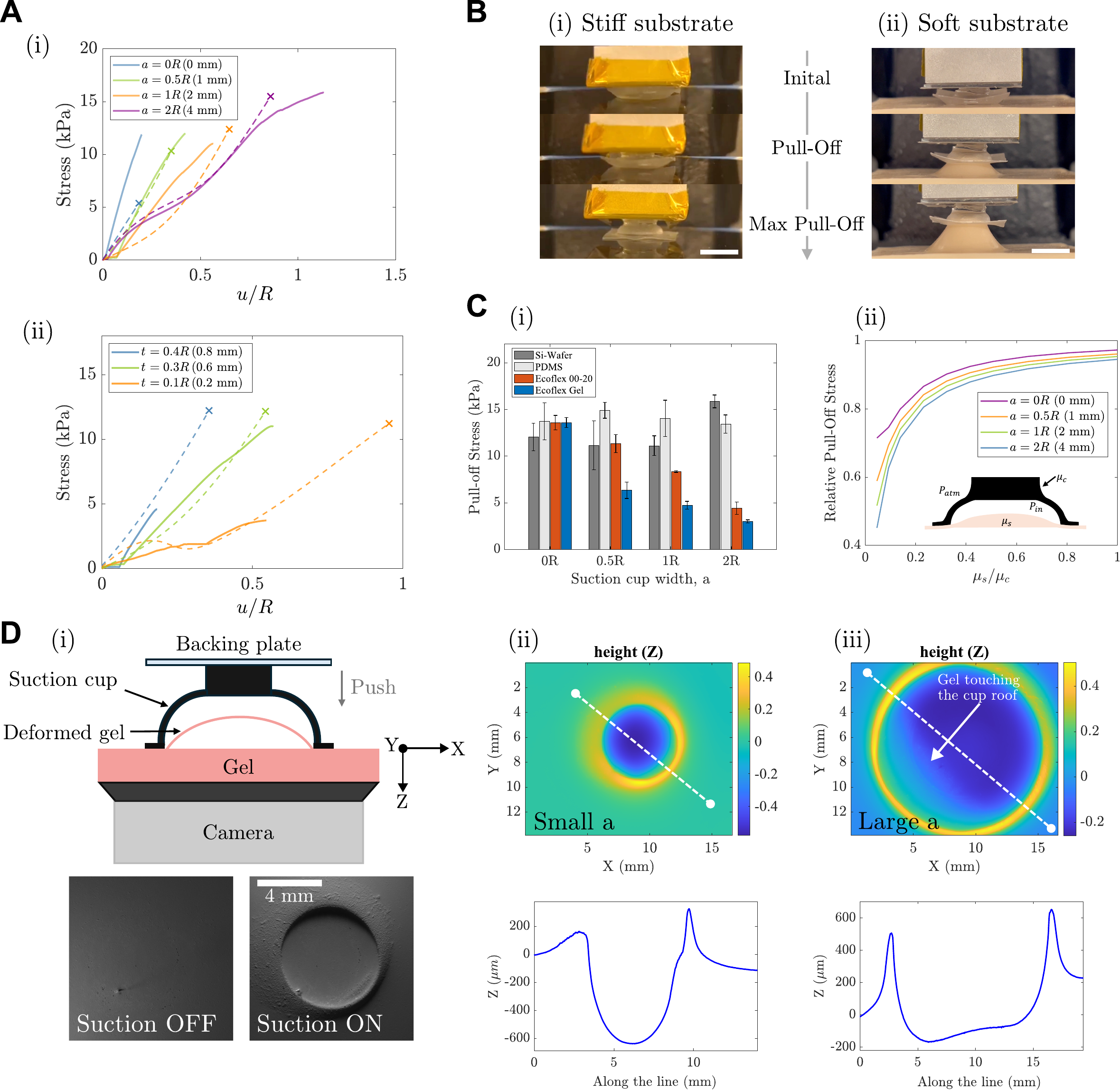}
\caption{\textbf{Effects of Target Surface Stiffness on Suction Performance.} \textbf{(\textit{A})} Normal pulling stress vs. normalized pulling displacement ($u/R$) of a suction cup on a stiff surface (Si wafer) with varying (i) cup width, a, and (ii) cup thickness ($t$). Wider and thicker cups in general exhibit stronger adhesion. Experimental trends (solid lines) align well with numerical modeling predictions (dashed lines). \textbf{(\textit{B})} Target surface stiffness significantly affects suction cup performance. Scale bar 10 mm. \textbf{(\textit{C})} (i) Normal adhesion strength of cups with varying a on surfaces with different stiffness (Si wafer = 100 GPa, PDMS = 1 MPa, Ecoflex 00-20 = 40 kPa, Ecoflex Gel = 20 kPa). (ii) Simulated relative pull-off stresses (normalized by the pull-off stress computed for the rigid substrate) for cups of varying radii on substrates of varying stiffness. \textbf{(\textit{D})} (i) Gelsight measurements indicate inward surface deformation upon suction. Scale bar 4 mm. Cup width, $a$ affects the degree of surface deformation relative to cup chamber volume: (ii) narrower cup induce smaller deformation, while (iii) wider cups show larger deformation, touching cup dome and occupying most of inner chamber, resulting in weaker vacuum.}
    \label{figure2}
    \index{figures}
\end{figure*}

We first evaluate the suction performance on rigid silicon wafers using cups with radius $R=$2 mm. Following the loading protocol in Fig.~\ref{figure1}\textit{D}, the cups are compressed by $\Delta=$1.25mm ($\varepsilon=\Delta/R = 0.625$), then rapidly unloaded to activate suction via elastic recovery. Once suction stabilizes, a normal pull test is performed at a rate of 2 mm/min (see Materials and Methods). As shown in Fig. \ref{figure2}\textit{A}, wider roof designs (larger $a$) produce higher pull-off stress $P_{pull-off}$, due to the increased undeformed cavity volume and thus a larger pressure drop during recovery. Similarly, thicker shells (higher $t$) generate stronger elastic recovery and higher suction forces (Fig.~\ref{figure2}\textit{A},ii). However, it is noted when thickness increases beyond a certain limit ($t/R\geq0.4$), friction at the contact interfaces is not strong enough to hold the cup rim in place. This results in local slip and air leakage in the deformation recovery phase, ultimately reducing suction strength. These trends are consistent between experiments (solid lines) and simulation results (dashed lines), validating our mechanical model.

As illustrated in Fig.~\ref{figure2}\textit{B}, substrate stiffness has a strong impact on the suction performance. On soft materials, the substrate's deformation under vacuum becomes comparable to the cup’s dimensions. This results in coupled cup–substrate interactions, where the dependency on cup design, particularly the cup aspect ratio $a/R$, becomes more complex as shown in Fig.~\ref{figure2}\textit{C}, i. For narrow cups ($a=0$), the performance remains relatively stable across substrates of varying stiffness, with $P_{pull-off}$ slightly higher on softer substrates. In contrast, wider cups with larger $a$ show a dramatically reduced pull-off strength as the stiffness of the substrate decreases. Specifically, for the $a=2R$ cups, the $P_{pull-off}$ on the rigid Si Wafer drops by approximately 15\%, 72\%, and 82\% when moving to PDMS, Ecoflex 00-20, and Ecoflex Gel, respectively. This is supported by the simulation results of Fig.~\ref{figure2}\textit{C}, ii, which shows that pull-off stress is increasingly sensitive to the substrate stiffness as the cup width $a$ increases. This demonstrates a clear inverse correlation between suction performance and substrate compliance for the flatter dome geometries (larger $a/R$).

While the dependency on roof width $a/R$ predominantly governs suction behavior, shell thickness $t$ also plays an important role across substrates of varying stiffness. Thicker shells generate stronger elastic recovery, resulting in higher $P_{pull-off}$, consistent with trends observed on rigid surfaces (see Fig.~\ref{figure2}\textit{A} ii). However, on highly compliant substrates (e.g., Ecoflex Gel), the influence of shell thickness becomes less pronounced, as moderate wall thickness is sufficient to recover most of the cup deformation (see \textit{SI Appendix, Fig. S5}). At higher thickness to radius ratios, increased shell stiffness hinders effective cup deformation. Hence, a greater proportion of the applied load is transferred to the underlying substrate directly through the cup structure instead of being used for cup deformation. This reduces suction efficiency and increases local stress at the skin-cup interface, which may be undesirable for wearable applications.

To directly visualize these effects, we employ a GelSight system to capture the deformation of the substrate inside the cup during suction activation (Fig.~\ref{figure2}\textit{D},i). The deformation profiles reveal a clear dependence on cup geometries. For cups with smaller roof widths (low $a/R$), substrate deformation remains localized. Here, only a limited portion of the cup cavity (approximately 15\% of the total volume) is filled with the deformed substrate, as shown in the cross-sectional image in Fig.~\ref{figure2}\textit{D},ii. In contrast, for high $a/R$ designs, the substrate is drawn deeply into the dome, making contact with the inner roof surface (Fig. \ref{figure2}\textit{D},iii). This extensive infill drastically reduces the recoverable volume and therefore the vacuum that can be generated (more data in \textit{SI Appendix, Section 5}). These results offer direct experimental evidence of the coupled deformation between soft substrates and suction cups, implying that suction cup designs optimized for rigid surfaces (e.g., flatter domes with wide roofs) are not effective for compliant skin-like substrates. Instead, low aspect ratio (small $a/R$) geometries such as those found in traditional cupping therapies are more suitable for maximizing adhesion on soft tissue. These designs preserve a larger recoverable volume while minimizing relative substrate intrusion during elastic rebound. 
For the effect of shell thickness, substrate deformation is similar for $t/R=0.2$ and $t/R=0.3$. However, significantly shallower deformation is observed at $t/R=0.4$ (see \textit{SI Appendix, Fig. S13}). These results experimentally verify that vacuum pressure generation decreases at higher shell thickness due to restricted cup deformation and increased risk of air leakage.

\subsection*{Anti-leaking Design for Skin Roughness}
Beyond compliance, a critical factor influencing the suction performance on skin is the surface roughness. Human skin features microtopographical wrinkles and irregularities, with roughness amplitudes ranging from sub-micron levels to tens of micrometers, depending on body location and other physiological factors~\cite{campolo_new_2010, derler_tribology_2012}. These microgaps pose a significant challenge for suction-based adhesion, as they can become leakage pathways that undermine sealing. Thus, it is essential to ensure a conformal, airtight interface between the suction cup and skin.

A commonly used strategy, inspired by biological systems (e.g. octopus suckers), is to use a thin liquid film that acts as a sealant at the contact interface. This approach can enhance pull-off stress by nearly an order of magnitude~\cite{wang_water_2022}. In our trials, the bare cup design functions effectively on moisture-rich regions, such as the fingertip and palm, where natural perspiration likely contributes to sealing~\cite{yue_bioinspired_2024}. However, in drier regions of the wrist, chest, or back, where hydration is limited, wet adhesion becomes unreliable. 

\begin{figure*}
\centering
    \includegraphics[width=0.9\textwidth]{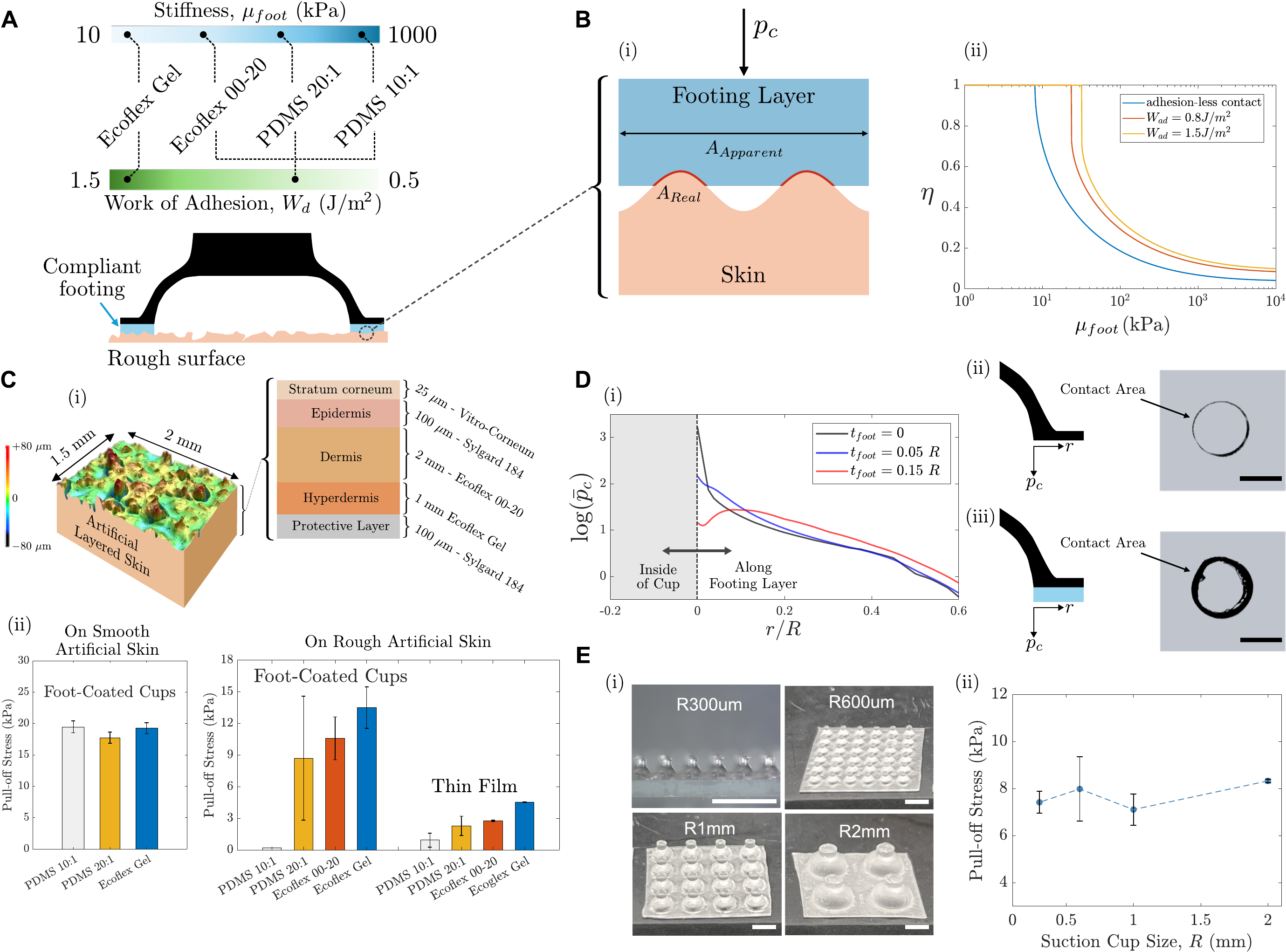}
    \caption{\textbf{Real Contact Area and Anti-Leakage Design for Suction Cups.} \textbf{(\textit{A})} Schematic of suction cups integrated with compliant footing layers of varying modulus and work of adhesion to enhance sealing on rough skin surfaces. \textbf{(\textit{B})} Analytical prediction of the real contact area ratio (real contact area divided by total footing area) based on adhesive contact theory. Skin undulation magnitude ($h=\pm80\mu m$), wavelength ($\lambda=400\mu m$), and the average contact pressure ($p_c=20$ kPa) are assumed for the estimation. Results highlight the effects of applied pressure, modulus, and interfacial work of adhesion. \textbf{(\textit{C})} (i) Surface roughness profile of the artificial skin measured by confocal microscopy. Pull-off stress on (ii) smooth, roughness-free skin replica and on rough artificial skin with different footing layer materials. The results reveal the role of compliance and adhesion in suction performance on rough surfaces. Thin-film controls are included for comparison.  \textbf{(\textit{D})} (i) Simulated contact pressure distribution at the cup–skin interface. Optical scan of contact area (ii) without and (iii) with a compliant footing layer, demonstrating improved conformability and more uniform pressure distribution. Scale bar 1 mm. 
    \textbf{(\textit{E})} (i) Suction cup arrays with varying cup radii ($R$) and fixed geometry ratios ($a = R$, $t = 0.3R$). (ii) Normal adhesion strength results tested on Ecoflex 00-20 substrates confirm size-invariant suction performance. Scale bar 4 mm.}
    \label{figure3}
    \index{figures}
\end{figure*}

To enable more consistent performance across a diverse set of skin sites, we adopt a design strategy that incorporates a soft, conformal interface layer between the suction cup and skin. This is inspired by previous demonstrations using soft interfacial coatings~\cite{yue_bioinspired_2024, lee_softened_2023}. The footing layer serves to fill surface microgaps and enhances sealing, as shown in Fig. \ref{figure3}\textit{A}. While such modifications are often implemented empirically, we take a systematic approach by establishing a quantitative analysis that links both interfacial compliance and work of adhesion to sealing effectiveness, enabling the design of optimized anti-leaking interfaces tailored for skin roughness.

To evaluate how interfacial compliance and adhesion contribute to sealing, we employ an analytical contact mechanics model tailored to skin-like surfaces. Existing air leakage models are often based on percolation theory, applying well to randomly rough but nominally flat surfaces~\cite{C9SM01679A}. However, human skin, particularly for the hand and wrist, exhibits more regular, periodic undulations, with a characteristic wavelength of approximately $\sim$0.4$mm$~\cite{Kovalev2014}. For such surfaces, leakage is primarily driven by a lack of conformal contact over these larger undulations, rather than microscopic asperities. Therefore, we model the skin surface as a simplified 1D sinusoidal profile with wavelength $\lambda$ and a small amplitude $h$. The suction cup’s soft footing layer is modeled as a nominally flat elastic solid. 

Following Johnson’s adhesive contact framework~\cite{johnson_adhesion_1995}, we capture the combined effects of materials' compliance, with equivalent modulus $E^*$, and adhesion, through the work of adhesion $W_{ad}$. $E^*$ is defined as $\frac{1}{E^*} = \frac{1-\nu_{foot}^2}{E_{foot}} + \frac{1-\nu_{skin}^2}{E_{skin}}$. Here, $E$ and $\nu$ are the Young’s modulus and Poisson’s ratio for each material. The work of adhesion is normalized by the pressure scale and skin undulation amplitude, $\overline{W}_{ad} = \frac{W_{ad}}{p^* h}$, where $p^*=\frac{\pi E^* h}{\lambda}$ is the pressure required to achieve conformal contact in the absence of adhesion. This formulation~\cite{johnson_adhesion_1995} allows us to relate the real contact area ratio $\eta=A_{Real}/A_{Apparent}$, which captures the degree of sealing at the interface, to the interface materials' equivalent modulus and adhesion under an applied average contact pressure $p_c$,
\begin{equation}
    \frac{p_c}{p^*}=\sin^2 \left(\frac{\pi}{2}\eta \right)-\left( \frac{2}{\pi} \ \overline{W}_{ad} \ \tan \left(\frac{\pi}{2}\eta \right)\right)^{1/2}.
\end{equation}

As plotted in Fig. \ref{figure3}\textit{B}, the model reveals that achieving conformal contact with effective sealing ($\eta = 1$) requires both low modulus and high adhesion. Interestingly, the relationship between $\eta$ and interface compliance $\mu_{foot}$ is highly nonlinear; the contact ratio $\eta$ remains low until the material stiffness falls below a critical threshold, beyond which $\eta$ increases sharply. This highlights the importance of selecting materials that are soft enough to cross this threshold. Moreover, increasing the work of adhesion has a two-fold benefit: it both elevates the overall contact ratio $\eta$ while also raising the critical stiffness threshold. This provides greater design flexibility and underscores the dual role of compliance and adhesion in mitigating air leakage, particularly for rough, low-hydration skin surfaces.

To validate our theoretical findings, we fabricate an artificial skin specimen that replicates both the compliance and surface texture of human skin. The underlying skin compliance was modeled by stacking three elastomer layers, each mimicking a distinct layer stiffness of the epidermis-dermis-hypodermis composite (see \textit{Materials and Methods}). To simulate realistic roughness, we laminate the stack with a commercially available stratum corneum mimic (Vitro-Skin\textsuperscript{\small{\textregistered}}), which exhibits an average surface roughness ($R_z$) of approximately $\pm$80 $\mu$m as measured by confocal microscopy (see Fig.~\ref{figure3}\textit{C}, i, and \textit{SI Appendix, Fig. S9}). Onto the foot region of the suction cups, we dip-coat four different interfacial materials with varying moduli from 1 MPa to 20 kPa and work of adhesion from 0.8 $J/m^2$ to 1.5 $J/m^2$ at approximately $\sim$200 $\mu$m thickness(Fig.~\ref{figure3}\textit{A}). The moduli and the work of adhesion of the interfacial materials are quantified using standard tensile and peeling tests (see \textit{SI Appendix, Fig. S14}).

As a control, we first test the footed cups on a smooth, roughness-free skin model (i.e., compliant bed only). Under these idealized conditions, all interface materials exhibit similar suction performance, indicating minimal influence from either compliance or adhesion (Fig.~\ref{figure3}\textit{C}, ii, left-panel). However, when tested on the rough skin replica, the effects of interfacial compliance and adhesion become prominent (Fig.~\ref{figure3}\textit{C}, ii, right-panel). For instance, PDMS 10:1 ($\mu_{foot} \approx$ 1 MPa, $W_{ad} \approx$ 0.89 $J/m^2$), the stiffest material tested, fails to maintain suction, likely due to poor conformability and persistent air leakage. Based on our contact model (Fig.~\ref{figure3}\textit{B}), the estimated contact ratio ($\eta$) for this condition remains below 0.2. A softer variant with similar $W_{ad}$, PDMS 20:1 ($\mu_{foot} \approx$ 400 kPa, $W_{ad} \approx$ 0.84 $J/m^2$), enhances suction strength by nearly 9 times. However, large standard deviations in performance suggest inconsistent sealing. This is likely a consequence of transitioning across a critical compliance threshold, where small improvements in contact can yield large gains, but are not yet sufficient for reliable attachment. In contrast, Ecoflex 00-20, a more compliant elastomer ($\mu_{foot} \approx$ 40 kPa) with comparable adhesion ($W_{ad} \approx$ 0.79 $J/m^2$), produces stronger and more consistent suction. The contact ratio $\eta$ exceeds 0.5, which is correlated with a noticeable improvement in sealing consistency and strength. Finally, Ecoflex Gel($\mu_{foot} \approx$ 20 kPa, $W_{ad} \approx$ 1.48 $J/m^2$) yields the best performance on rough skin, achieving robust and repeatable sealing. Despite having a similar modulus to Ecoflex 00-20, its higher work of adhesion enables near-complete conformal contact ($\eta \approx 1$). This result aligns with our model’s prediction of the synergistic role of compliance and adhesion in improving both contact ratio and sealing robustness.

To isolate the contributions of suction versus material adhesion, we compare suction cups against flat films of the same materials. In all cases except PDMS 10:1, the footed suction cups outperform the films, demonstrating that suction significantly enhances adhesion, but only when a proper seal is maintained. Once leakage occurs, suction becomes inactive and performance can fall below the adhesion of the bare material.

Beyond improving sealing, the compliant footing layer also modulates the stress distribution at the skin–cup interface. Finite element simulations reveal that increasing the footing layer thickness redistributes the contact pressure more uniformly across a broader area (Fig.~\ref{figure3}\textit{D}, i) in the equilibrated, unloaded state. Here, an initial indentation of $\Delta/R = 0.5$ is applied for all of the cases. The contact pressure is normalized by the nominal contact pressure, that is, the total contact force divided by the entire area of the footing layer. In the absence of a footing layer, stresses concentrate sharply along the inner rim of the suction chamber, resulting in localized pressure peaks. A soft, compliant footing layer attenuates these peaks and distributes the load more evenly across the contact area. 

We support these simulation results experimentally using an optical contact mapping setup (see \textit{SI Appendix, Section 7}). Without a footing layer, sealing is confined to a narrow circumferential ring (Fig.~\ref{figure3}\textit{D}, ii). In contrast, using Ecoflex Gel as a footing material significantly increases the real contact area, enabling broader surface engagement (Fig.~\ref{figure3}\textit{D}, iii). This enlarged contact area reduces stochastic air leakage pathways and improves user comfort by minimizing localized pressure points. When tested on real human skin, it was qualitatively reported that the footed vacuum cups maintain stable attachment for several hours without loss of pull-off strength or signs of user discomfort.

Having optimized suction performance through geometric and material parameters, we next evaluate whether these principles hold across different size regimes. By scaling the cup radius $R$ from 2 mm down to 300 $\mu$m while maintaining constant geometric ratios ($a/R=1$, $t/R=0.3$), we find that both the vacuum level and pull-off strength remain largely unchanged (Fig.~\ref{figure3}\textit{E}). This size-invariant behavior is observed across both compliant substrates and rough artificial skin (see \textit{SI Appendix, Fig. S6}), indicating that the suction mechanism operates independently of absolute size when the relative geometry is preserved. These results suggest that the design principles established at the millimeter scale can be extended to smaller dimensions, enabling broad applicability across different skin sites and device sizes.

\subsection*{Transforming Conventional Electronics into Wearables}
\begin{figure*}
\centering
    \includegraphics[width=0.9\textwidth]{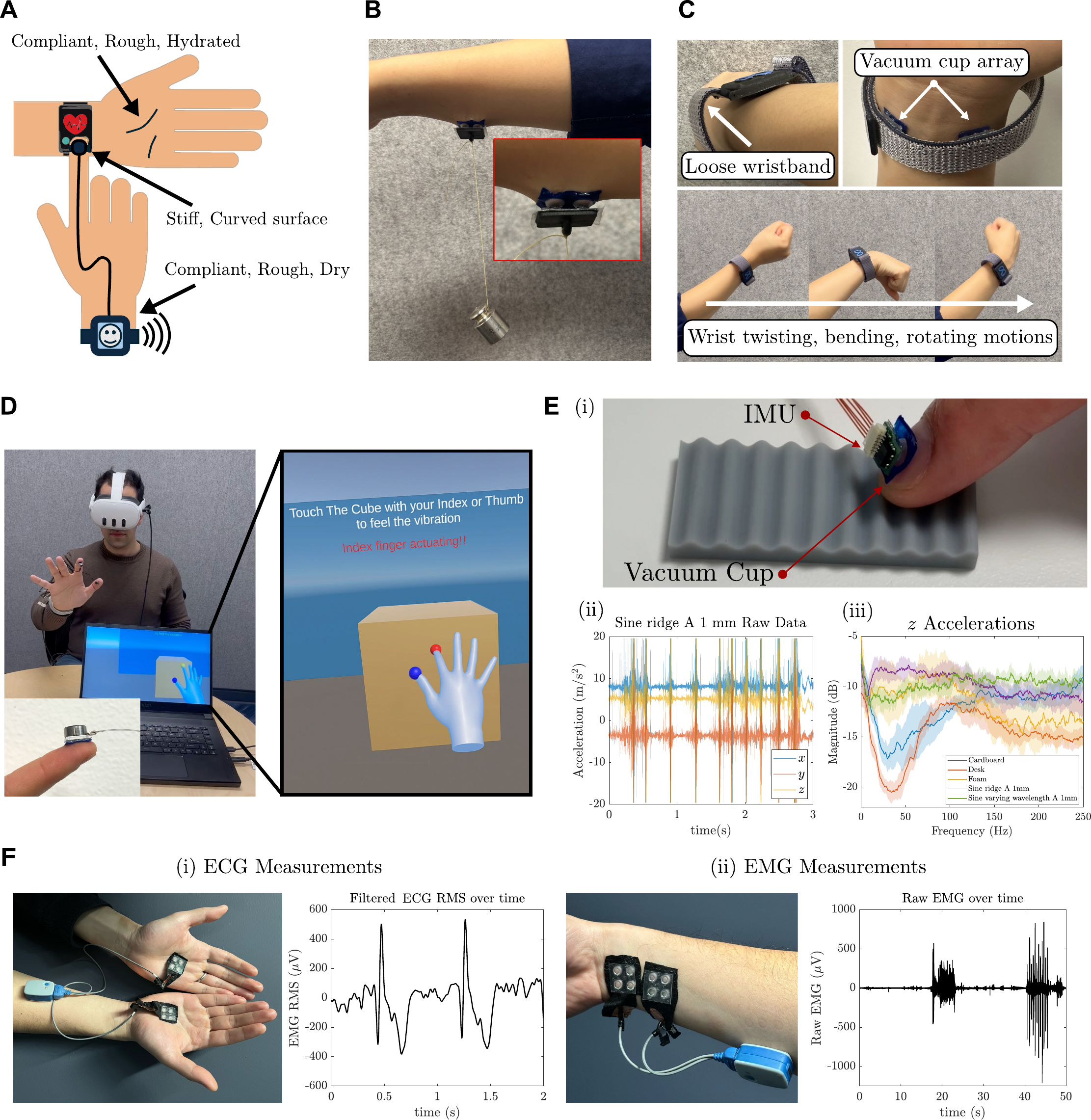}
    \caption{\textbf{Demonstration of vacuum cup applications for wearable electronics and skin adhesion.} \textbf{(\textit{A})} A schematic showing the need for securing electronics attachment across various hand regions and skin conditions. \textbf{(\textit{B})} A 2$\times$2 suction cup array integrated onto a rigid acrylic backing supports a 100 g weight on the forearm, demonstrating strong normal adhesion to compliant skin regardless of backing stiffness. Inset: close-up showing large skin deformation and firm vacuum cup attachment. \textbf{(\textit{C})} A wearable wristband integrated with vacuum cups enables secure, on-demand skin adhesion without requiring tight compression. Even under loose fit and dynamic motion, the suction interface prevents sliding and detachment. \textbf{(\textit{D})} Suction-mounted Linear Resonant Actuators (LRAs) on the thumb and index finger deliver localized haptic feedback in a VR environment. \textbf{(\textit{E})} (i) Inertial Measurement Units (IMUs) attached to fingernails via suction cups enable motion tracking and surface texture detection. Representative (ii) time- and (iii) frequency-domain signals highlight stable contact and high signal differentiability during sliding interactions with various substrates. \textbf{(\textit{F})} Electrophysiological signals for (i) ECG and (ii) EMG measured using suction cups with conductive footing layers (Ecoflex Gel infused with 20 wt\% carbon black). The modified cups maintain conformal skin contact while ensuring electrical connectivity.}
    \label{figure4}
    \index{figures}
\end{figure*}

Building on the fundamental insights into suction mechanics with skin-like substrates, we demonstrate how our deformation-driven vacuum cups can transform both rigid and soft conventional electronic components into skin-interfacing, wearable devices. This approach enables robust, reliable attachment to skin, offering a versatile interface for user-device interactions across a diverse set of anatomical sites and mechanical contexts Fig.~\ref{figure4}\textit{A}.

We first evaluate the weight-bearing capability of the suction cup array. As shown in Fig.~\ref{figure4}\textit{B}, a 2$\times$2 array integrated onto a rigid acrylic backing successfully supports a 100 g load when adhered to a user’s forearm. This confirms the ability of the vacuum cups to sustain significant normal forces on soft, compliant skin, independent of the mechanical stiffness of the back layer. Leveraging the design principles developed earlier for different substrate stiffness, we engineer a double-sided vacuum cup array capable of interfacing simultaneously with substrates of disparate stiffness. As demonstrated in Movie S1, one side of the device employs flat suction cups optimized for rigid surfaces (e.g., electronics or weights), while the opposing side features narrower cups with low $a/R$ ratios, tailored for soft, deformable skin-like substrates. Reversing the array orientation such that each side is misaligned with its intended substrate significantly degrades performance. This reinforces the importance of a substrate-specific suction design methodology.

The reliable performance of suction cups may help resolve the trade-off between skin-device contact quality and user comfort. For wearable electronic devices such as smart watches, loose bands may have frequent contact losses between biosensors and skin, while tight, compressive bands often cause discomfort during prolonged use. By integrating our carefully designed suction cups with a conventional wristband, we achieve strong, stable skin adhesion without relying on restrictive fits. This configuration supports on-demand attachment, allowing users to press and engage the vacuum cups as needed. As shown in Fig.~\ref{figure4}\textit{C} and Movie S4, even with a loosely worn band, the suction interface prevents sliding and detachment under motion. 

The suction adhesives also enable direct integration of functional electronics for sensing and actuation. By selecting appropriate suction cup geometries, devices can be securely attached to regions of the body with different curvatures and stiffness, including the fingertip, palm, nail, and knuckle. In Fig.~\ref{figure4}\textit{D and E} (and Movie S2 and S3), we demonstrate reliable attachment of inertial measurement units (IMUs) and linear resonant actuators (LRAs) to the small, mobile skin areas of the fingernail and fingertip. The attached IMUs maintain stable contact during finger motion, enabling accurate measurement of frequency signals generated when sliding across surfaces with varying textures (Movie S2). This stable interface minimizes relative displacement and reduces signal noise for micro-vibration detection that could support surface texture classification. Similarly, the LRAs attached to the user's index finger and thumb successfully deliver localized haptic feedback without restricting natural hand movement. When the user interacts with virtual objects, contact events trigger immediate vibration through the LRAs, creating a realistic tactile response during interaction. The suction cups maintain stable adhesion throughout actuation, and we observe no degradation or mechanical failure even under continuous vibration at 130 Hz across $>$ 20 don-doffs tested. 

Finally, we extend the platform to enable direct skin contact for electrophysiological sensing, and demonstrate it for ECG and EMG monitoring. By doping the suction cup’s soft footing layer (Ecoflex Gel) with carbon black, we create a conductive, compliant interface that maintains both suction integrity and electrical connectivity. As demonstrated in Fig.~\ref{figure4}\textit{F} and Movie S5, the modified suction cups enable high-fidelity biosignal acquisition using a commercial monitoring system (Ultium EMG, Noraxon USA Inc.). In a dual-wrist ECG configuration, the system reliably captures clear QRS waveforms under static conditions. For EMG, bipolar electrodes placed across the wrist and forearm successfully record muscle activity during voluntary contraction and finger pinch and release motions (Movie S5). This highlights the platform’s potential for soft, skin-conformal health monitoring systems that combine mechanical stability with robust electrical performance.

%%%%%%%%%%-------------------Discussions-------------------%%%%%%%%%
\section*{Discussion}
In this work, we present a deformation-driven vacuum cup array that enables robust, reliable adhesion to human skin, offering a novel, power-free platform for transforming conventional electronic components into skin-wearable devices. By systematically investigating suction mechanics on soft and rough substrates, we establish design principles that guide the geometry, compliance, and adhesion tuning necessary for robust, stable attachment across a variety of skin conditions. These insights allow us to demonstrate versatile integration of sensing and actuation components, including inertial sensors, haptic actuators, and electrophysiological electrodes without the need for tight bands, adhesives, or bulky external supports.

While the demonstrated approach shows strong potential, several design and fabrication challenges remain to be solved. The current suction cup geometries, inspired by traditional cupping therapies, generate high vacuum pressure but are not necessarily optimized for mechanical efficiency or elastic recovery. Future work may benefit from computational shape design and optimization~\cite{bendsoe1995optimization,Akerson2023,Akerson2022}, or bioinspired suction geometries~\cite{Yuewang2022} to improve force generation, conformability, and reusability. In parallel, advancements in microfabrication techniques, such as multi-photon polymerization~\cite{skliutas_multiphoton_2025}, micro/nano patterning~\cite{zhang_patterning_2010, kagias_metasurface-enabled_2023} or silicon-based etching~\cite{baik_wet-tolerant_2017, choi_cephalopodinspired_2016, choi_colloidal_2004}, could enable further miniaturization down to sub-100 $\mu$m scales. Such capabilities would expand the platform to applications in microscale electronics, invisible skin interfaces, and soft microrobotics, where thinner, more discreet adhesion layers are required.

Material advancements will also play a key role in the system’s long-term usability. While ultra-soft, tacky footing layers enhance sealing on rough or dry skin, they are susceptible to contamination, abrasion, and degradation~\cite{liu_silicone-based_2017}. The use of encapsulation with smart coatings such as self-healing, self-cleaning or functional elastomers~\cite{khan_self-healing_2020, menguc_staying_2014, zarei_advances_2023, jiang_abrasion_2021, lee_molecular_2024} may help maintain both high adhesion and mechanical resilience over repeated use cycles and environmental exposure.

Beyond the design space explored in this study, additional physiological and mechanical features of real skin must be considered. For example, hairy skin surfaces introduce gaps that may compromise sealing performance. Downsizing the suction cup radius to the sub-100 $\mu$m scale could allow individual cups to fit between hair strands and form localized seals~\cite{baik_highly_2018}. Similarly, skin curvature and dynamic undulations, arising from muscle movement, joints, or vascular structures, introduce challenges for maintaining continuous contact. Optimizing cup size, array layout, and backing layer compliance is critical to ensure conformability and reliable adhesion under such conditions. For instance, smaller cups may require highly flexible substrates to conform over curved surfaces, while larger cups can individually deform to accommodate curvature, even when integrated with rigid electronics. As demonstrated in our example of adhering an IMU to the fingernail, larger cups are more effective than dense arrays of smaller cups on a rigid backing, which fail to fully engage with curved regions.

Looking ahead, wearable electronics are integral to achieving the vision of human computer symbiosis~\cite{licklider_man-computer_1960}. Stable attachment of these devices may allow for sensing of a users’ intentions, the environment, and the context around them. The ability to transform conventional electronics into wearables without custom mechanical packaging or power-dependent adhesion opens new possibilities for human-device interaction. Our platform provides a scalable, low-barrier pathway to prototype and deploy skin-mounted systems in fields such as health monitoring, assistive technologies, and immersive interfaces. Beyond skin, our modeling and material insights apply more broadly to adhesion on other rough, soft, or dynamic surfaces. It suggests opportunities in soft robotics, bioadhesives, and adaptive interfaces for unstructured environments.

%%%%-----SI (please check SI folder for actual document----%%%%%%%
\newpage
\section*{\large Supporting Information Appendix (SI)}
All study data are included in the article and/or \textit{SI Appendix}.

% \subsection*{SI Movies}
% Supply Audio Video Interleave (avi), Quicktime (mov), Windows Media (wmv), animated GIF (gif), or MPEG files. Movie legends should be included in the SI Appendix file. All movies should be submitted at the desired reproduction size and length. Movies should be no more than 10MB in size.

%%%%%%%%%%-----------------Materials and Methods--------------%%%%%%%%%
\section*{Materials and Methods}
\subsection*{Mold Fabrication via Stereolithography}{
Two-part molds for fabricating the suction cups are produced using stereolithography (SLA) 3D printing. For cup radii $R>600\mu m$, we use a Formlabs Form 3B printer with Grey Pro Resin, while for smaller cups with 
$R=300\mu m$, high-resolution printing is performed using ProtoLabs’ MicroFine$^{\text{TM}}$ resin. After printing, both top and bottom mold components undergo standard post-processing specific to each resin system. This includes immersion in an isopropanol (IPA) bath for 15 minutes to remove uncured resin, followed by air drying. To achieve optimal mechanical integrity, the parts are then post-cured at 80\textdegree C for 15 minutes using a UV-curing unit. Due to the well-documented inhibition of silicone elastomer curing on SLA-printed surfaces from leaching of unreacted monomers and photoinitiators, an additional surface treatment is applied to mitigate curing inhibition. Specifically, molds are subjected to 8 hours of continuous 405 nm UV exposure, followed by a thermal bake at 70\textdegree C overnight. 
}

\subsection*{Vacuum Cups Fabrication}{
The main body of the vacuum cup array is fabricated from Sylgard 184 silicone elastomer (Dow Sylgard$^{\text{TM}}$) using a two-part molding process (\textit{SI Appendix, Fig. S1}). Sylgard 184 Part A (30 g) and Part B (3 g) are mixed at a 10:1 weight ratio using a planetary centrifugal mixer to ensure homogeneous mixing and degassing. The mixture is then poured into the assembled two-part mold through designated inlet channels. To eliminate residual air bubbles trapped in the mold cavities, the mold is placed in a vacuum chamber for 15 minutes. Following degassing, the sample is thermally cured in an oven at 70\textdegree C for 1 hour. Once cured and demolded, the vacuum cup array body is positioned onto a spin-coated layer of uncured Sylgard 184 (10:1 ratio) to form the back-layer assembly. After curing, this thin back layer is laminated onto a laser-cut acrylic sheet (1 mm thickness) using a double-sided adhesive film with a silicone adhesive on one side and acrylic adhesive on the other. This ensures uniform load transfer across the entire array during testing and application. To form the footing layer, we apply four different elastomeric materials: Sylgard 184 (10:1), Sylgard 184 (20:1), Ecoflex 00-20, and Ecoflex Gel (Smooth-On$^{\text{TM}}$). Each material is prepared as a 200 µm-thick film by spin-coating onto 51 mm × 51 mm glass slides. Sylgard elastomers are spin-coated at 1000 rpm, and Ecoflex elastomers at 3000 rpm, each for 30 seconds. The cured vacuum cup arrays are gently dip-coated onto the uncured elastomer films. The high viscosity of the films ensures that a clean and uniform transfer of the footing material onto the base of each cup is achieved. The footed assemblies are then thermally cured at 70\textdegree C for 15 minutes and then gently peeled off. For conductive applications, we fabricate a conductive footing layer by mixing Ecoflex Gel with 20 wt\% carbon black and spin-coating the mixture onto a PTFE substrate to form a thin film. The cured vacuum cup array (with back-layer attached) is laminated onto this uncured conductive layer and thermally cured at 80\textdegree C. Due to reduced tackiness caused by the filler, the conductive cups require manual trimming at the cup openings after peeling using scissors to expose the cup cavity.
}

\subsection*{Normal Pull-Off Stress Measurement}{
Normal pull-off stress is measured using a universal testing machine (Instron 5944, 50 N load cell), following a T-peel-like adhesion setup adapted from the ASTM F2258 standard (\textit{SI Appendix, Fig. S3}). Vacuum cups are mounted on a rigid acrylic backing and attached to the top compression clamp using a silicone–acrylic double-sided adhesive. Target substrates are affixed to the bottom compression clamp using the same adhesive. To activate suction, the sample is compressed to a displacement of $u/R=$0.625 at a rate of 10 mm/min. The system is then unloaded to 0 N rapidly, allowing elastic recovery and vacuum formation. After holding at 0 N for 30 seconds to reach equilibrium, the sample is pulled upward at 2 mm/min until detachment. The maximum tensile force recorded prior to failure is defined as the pull-off force, and the pull-off stress is calculated by dividing this value by the projected contact area of the cup. Tested substrates include silicon wafers, Sylgard 184 (10:1), Ecoflex 00-20, Ecoflex Gel, and artificial skin. All tests are conducted under ambient conditions, with at least three replicates performed for each condition.
}

\subsection*{Deformation Modeling}{
The full details on modeling formulation and numerics are presented in~\textit{SI Appendix, Section 3}. In summary, we assume an axi-symmetric domain and deformation, with all loading conducted in a displacement control setting. The initial indentation process assumes that the pressure inside and outside the cup are equal. Hence, we only consider the cup deformation with no contributions from the pressure differential across the cavity wall. We model this through finite deformation kinematics with a hyperelastic, weakly compressible Neo-Hookean material model. The equilibrium condition is found through minimizing the total system energy, and this is solved in a finite element settings. At the end of the loading process, the volume left inside the vacuum cup, $V_{0}$, sets the total amount of air during the recovery and pull-off process. The energy is modified to account for the work done by the assumed ideal gas. Minimization of this energy gives the equilibrium relations, and this is again solved in a finite element setting. The pull-off condition is determined when the total contact force on the bottom foot of the cup become tensile, that is, when a downward force is needed to keep the cup attached to the substrate. 
} 

\subsection*{Gelsight Suction Measurement}{
Substrate deformation under suction is evaluated using a Gelsight optical tactile sensor, which captures surface topology through deformation of an elastomer layer imaged by an integrated camera. A constant compressive strain ($\varepsilon=u/R=0.625$) is applied using a custom displacement-controlled actuator. The recorded images are reconstructed into 3D surface profiles, and cross-sections are analyzed to quantify deformation. Additional details are provided in \textit{SI Appendix, Fig. S10}.
}

\subsection*{Real Contact Area Measurement}{
To evaluate the real contact area of change due to the footing layer, a customized optical setup utilizing total internal reflection at contact interface is designed. The details on the setup are presented in \textit{SI Appendix, Section 7}.
}

\subsection*{Artificial Skin Replica Fabrication}{
The multilayer artificial skin replica is constructed to replicate the structural and mechanical properties of human forearm skin. The topmost stratum corneum layer consists of a commercial artificial skin film (Vitro-Corneum, VitroSkin Inc.) that mimics the surface texture and barrier properties of human skin. Beneath it, epidermis is modeled using a 100 $\mu$m-thick Sylgard 184 (10:1) replica containing skin-like surface topography. The dermis layer is cast from Ecoflex 00-20 (Smooth-On$^{\text{TM}}$), with a thickness of 2 mm and a nominal modulus of $\sim$40 kPa. A softer hypodermis layer is formed beneath using 1 mm of Ecoflex Gel (Smooth-On$^{\text{TM}}$), approximating the softness of subcutaneous tissue (Shore hardness 000). To provide structural support and facilitate handling, a 100 $\mu$m-thick Sylgard 184 film is laminated at the base as a protective backing layer.}

\subsection*{Artificial Skin Surface Roughness Measurement}{
Surface roughness of the artificial skin replica (Vitro-Skin\textsuperscript{\small{\textregistered}}) is characterized using a Keyence VK-X series confocal laser scanning microscope. Measurements are performed in standard vertical scanning mode with a 10$\times$ objective lens, covering a 1.5 mm $\times$ 2 mm area. The surface height map is reconstructed from multiple z-slices, and roughness parameters are extracted using Keyence MultiFileAnalyzer software. The average surface roughness ($R_Z$) is determined to be approximately 80 $\pm$ 5 $\mu$m, consistent with physiological skin topography. All samples are measured in ambient conditions without additional coating or surface treatment. The confocal microscope scan and roughness measurement of the artificial skin is shown in \textit{SI Appendix, Fig. S9}. 
}

\subsection*{LRA Demonstration with VR Integration}{
Two commercially available Linear Resonant Actuators (LRAs, VG0840001D, Vybronics Inc) are attached to the user's index finger and thumb using vacuum cup arrays  ($R=$300 $\mu$m, $a/R=$0.5, $t/R=$0.3). A haptics driver is used to drive the LRAs when triggered. To demonstrate realistic interaction, the device is integrated into a virtual reality (VR) environment. When users interact with virtual objects (e.g., a floating cube), contact-triggered haptic feedback is delivered through the LRAs. Additional implementation details are provided in the \textit{SI Appendix, Section 9}.
}

\subsection*{IMU Demonstration}{
An IMU (IIS3DWB, STMicroelectronics) together with a custom PCB for wire connection is attached to the fingernail of the user's index finger using a single vacuum cup ($R=$1 mm, $a/R=$1, $t/R=$0.3). The user then slide the finger against textured surfaces and the acceleration data in all 3 directions are recorded for analysis.
}

\subsection*{ECG/EMG Demonstration}{
To demonstrate biosignal sensing, 2$\times$2 vacuum cup arrays ($R=$1 mm, $a/R=$1, $t/R=$0.3) with a conductive footing layer are used as skin-conformal electrodes for ECG and EMG recording. A commercial wireless biosignal acquisition system (Ultium EMG, Noraxon USA Inc.) with a 24-bit ADC and 300 nV resolution is used for data collection at a 4000 Hz sampling rate. ECG measurements follow a dual-wrist configuration, with electrodes placed on both wrists and signals filtered between 0.05–300 Hz. For EMG, electrodes are placed on the forearm and wrist in a bipolar configuration, and signal envelopes are extracted using RMS with a 500 ms moving window and bandpass filtering (10–1000 Hz). Further experimental details and setup schematics are provided in the \textit{SI Appendix, Section 8}.
}

% \showmatmethods{} % Display the Materials and Methods section

%%%%%%%%%%-----------------Acknowledgements--------------%%%%%%%%%
\paragraph{Author Contributions }{S.L., A.A., D.P., C.D., and T.L. designed research;  S.L., A.A., R.P., E.H., K.R., Z.L., A.S., A.A., and T.L.,performed research;  S.L., A.A., R.P., E.H., K.R., Z.L., A.S., A.A., and T.L. analyzed data; and S.L., A.A., and T.L. wrote the manuscript.}

\paragraph{Acknowledgments}{All work was funded internally by Reality Labs Research, Meta Platforms Inc.. We thank Dr. Vanessa Tolosa for her contribution to project ideation. We also acknowledge Duane Irish for equipment support and technical assistance in this research. Useful discussions with Dr. Li Yao on conductive polymers are also gratefully acknowledged.
}
 % Display the acknowledgments section

%%%%%%%%%%-----------------Bibliography--------------%%%%%%%%%
% \bibsplit[30]
%Use \bibsplit to split the references from the body of the text. Value "[2]" represents the number of reference in the left column (Note: Please avoid single column figures & tables on this page.)

% Bibliography
% {\scriptsize
% % \input{Manuscript_arxiv.bbl} 
% \bibliographystyle{unsrtnat}     % or unsrt/ieeetr/plainnat/...
% \bibliography{Manuscript_arxiv}  % points to your .bib (no extension)
% }
% Bibliography
% % {\scriptsize
% \printbibliography}

\end{document}

%% Comment out or remove this line before generating final copy for submission; this will also remove the warning re: "Consecutive odd pages found".
% \instructionspage  
\maketitle

%% Adds the main heading for the SI text. Comment out this line if you do not have any supporting information text.

% \SItext
----------------------------------------------------------
%%% Each figure should be on its own page
\section*{1. Suction cup fabrication}
\begin{figure}[htb!]
    \centering
\includegraphics[width=1\textwidth]{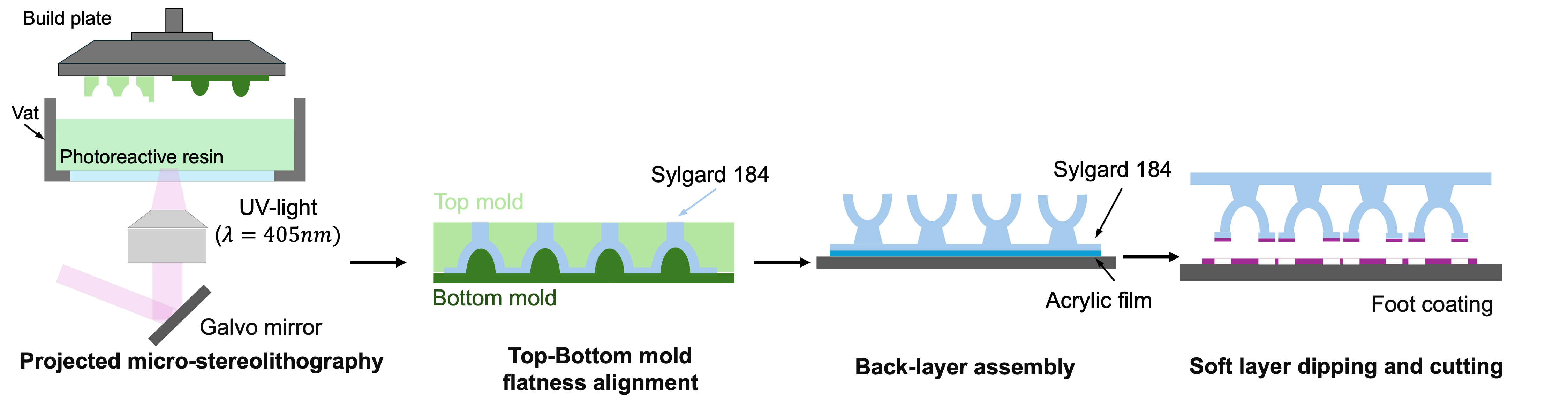}
    \caption{Suction cup fabrication processes.}
    \label{SIfig:fabrication_process}
\end{figure}

\begin{figure}[htb!]
    \centering
\includegraphics[width=0.9\textwidth]{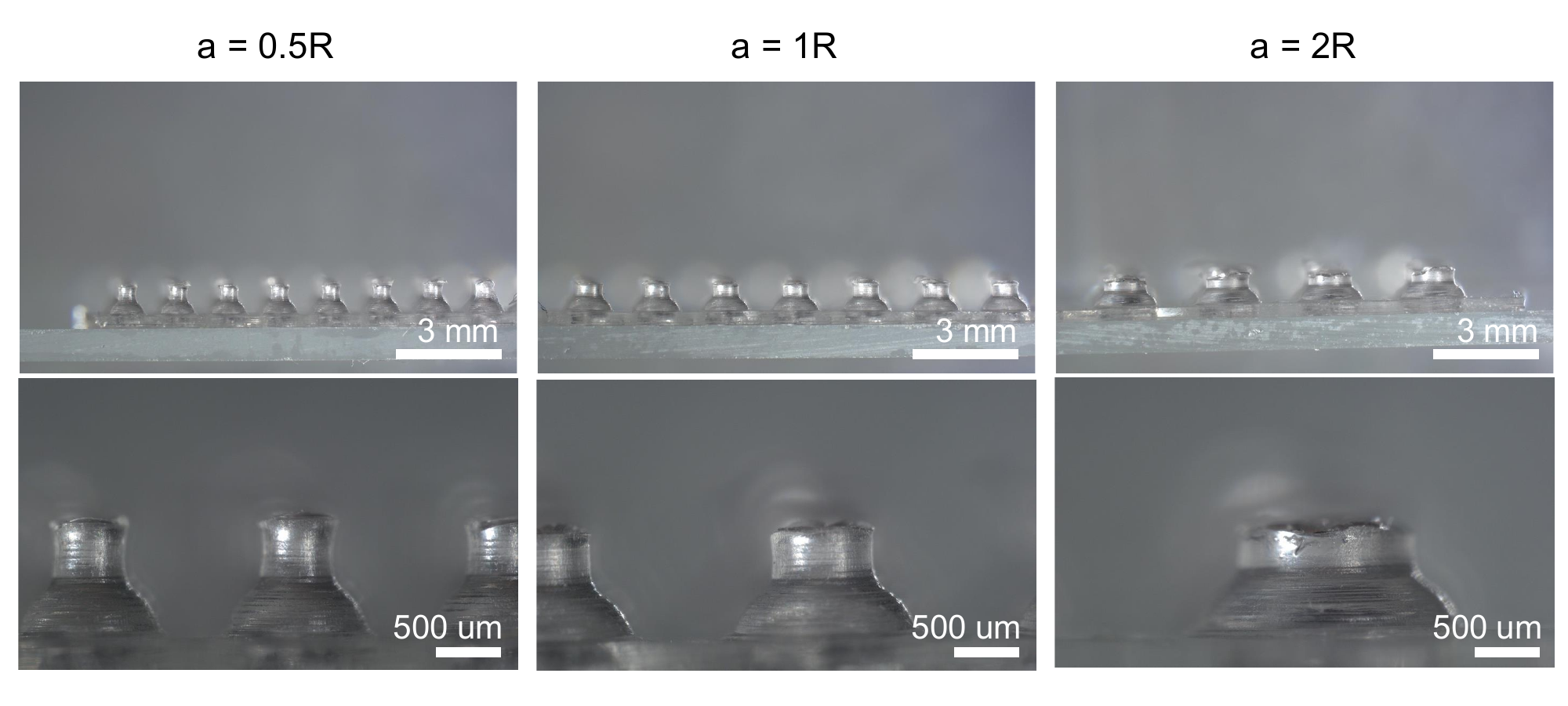}
    \caption{Optical images and measurements for main bodies of the cups with $R = 300 \mu m$ cups with varied $a$}
    \label{SIfig:microscop_a}
\end{figure}

\clearpage\section*{2. Normal pull-off measurements}
\begin{figure}[htb!]
    \centering
\includegraphics[width=0.85\textwidth]{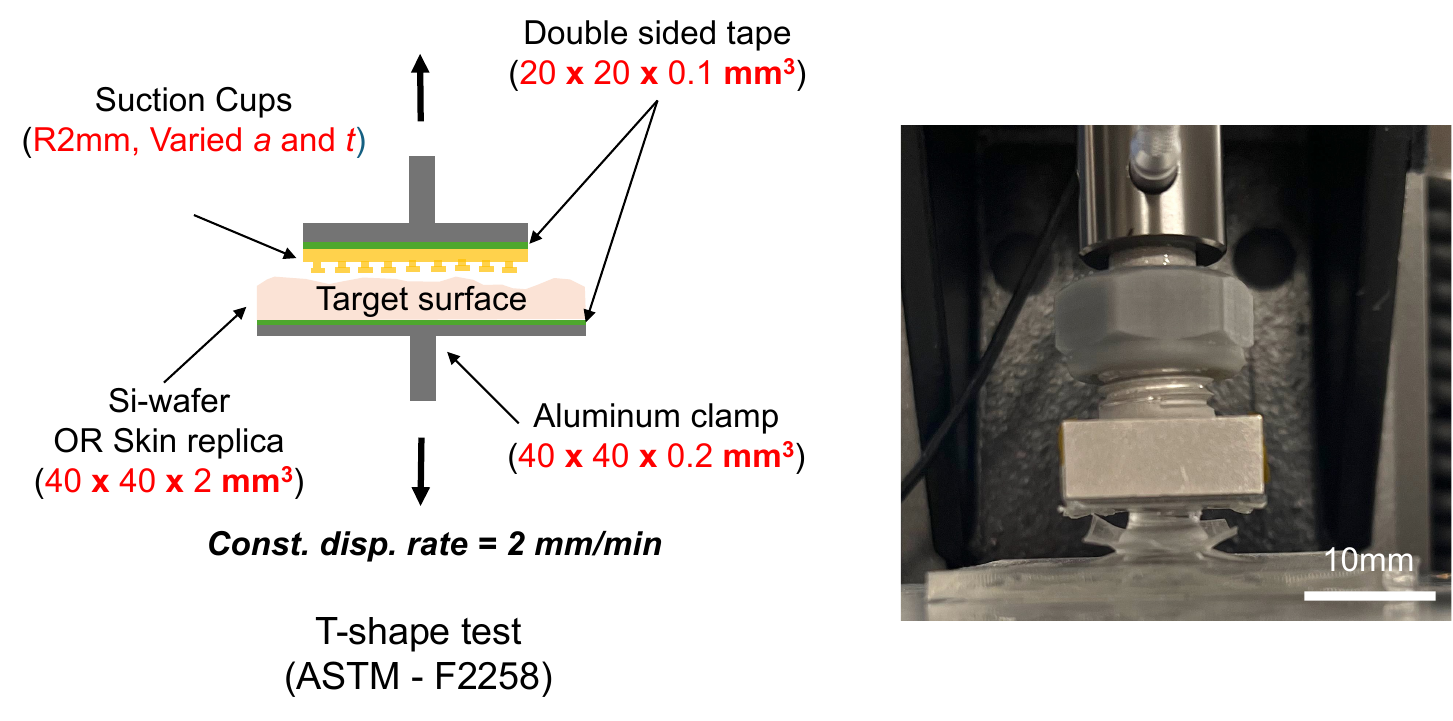}
    \caption{Pull-off strength measurement set-up}
    \label{SIfig:normal pull-off setup}
\end{figure}

\begin{figure}[htb!]
    \centering
\includegraphics[width=0.65\textwidth]{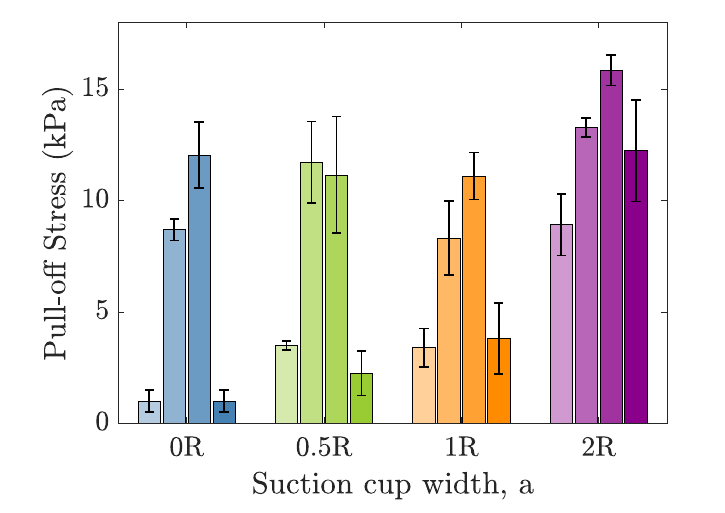}
    \caption{Histogram of pull-off strength for suction cups with varying roof width ($a$) and shell thickness ($t$) on a silicon wafer. Bar color indicates thickness: from lightest to darkest,  $t=0.1R$, $0.2R$ ,$0.3R$, $0.4R$ for $R=2mm$.}
    \label{SIfig:normal histogram_varied_t}
\end{figure}

\begin{figure}[htb!]
    \centering
\includegraphics[width=0.65\textwidth]{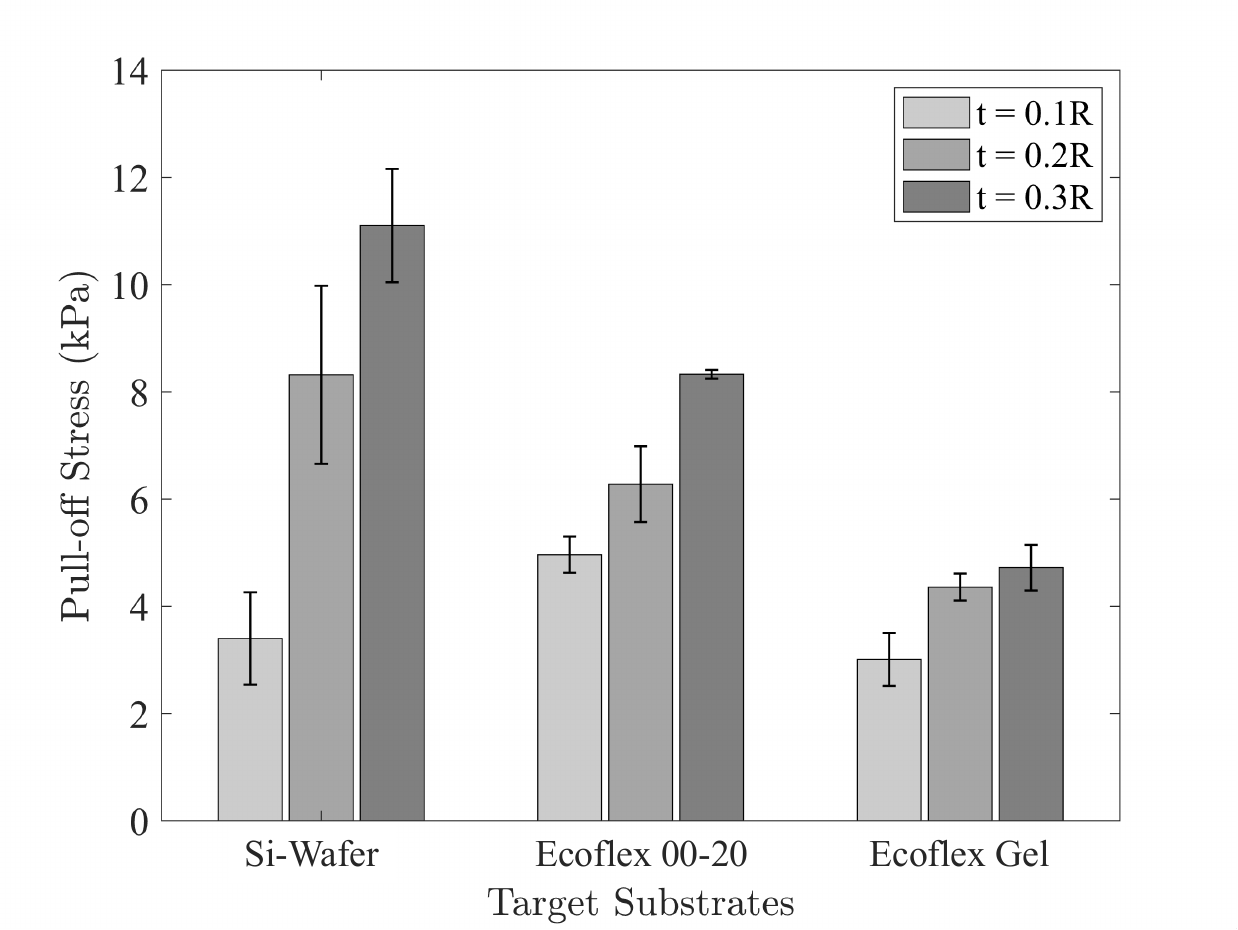}
    \caption{Histogram of pull-off strength for suction cups with varying cup thickness ($t$) on different target stiffness. Bar color indicates thickness: from lightest to darkest,  $t=0.1R$, $0.2R$ ,$0.3R$ for $a=R$ and $R=2mm$.}
    \label{SIfig:normal histogram_R2a2}
\end{figure}

\begin{figure}[htb!]
    \centering
\includegraphics[width=0.7\textwidth]{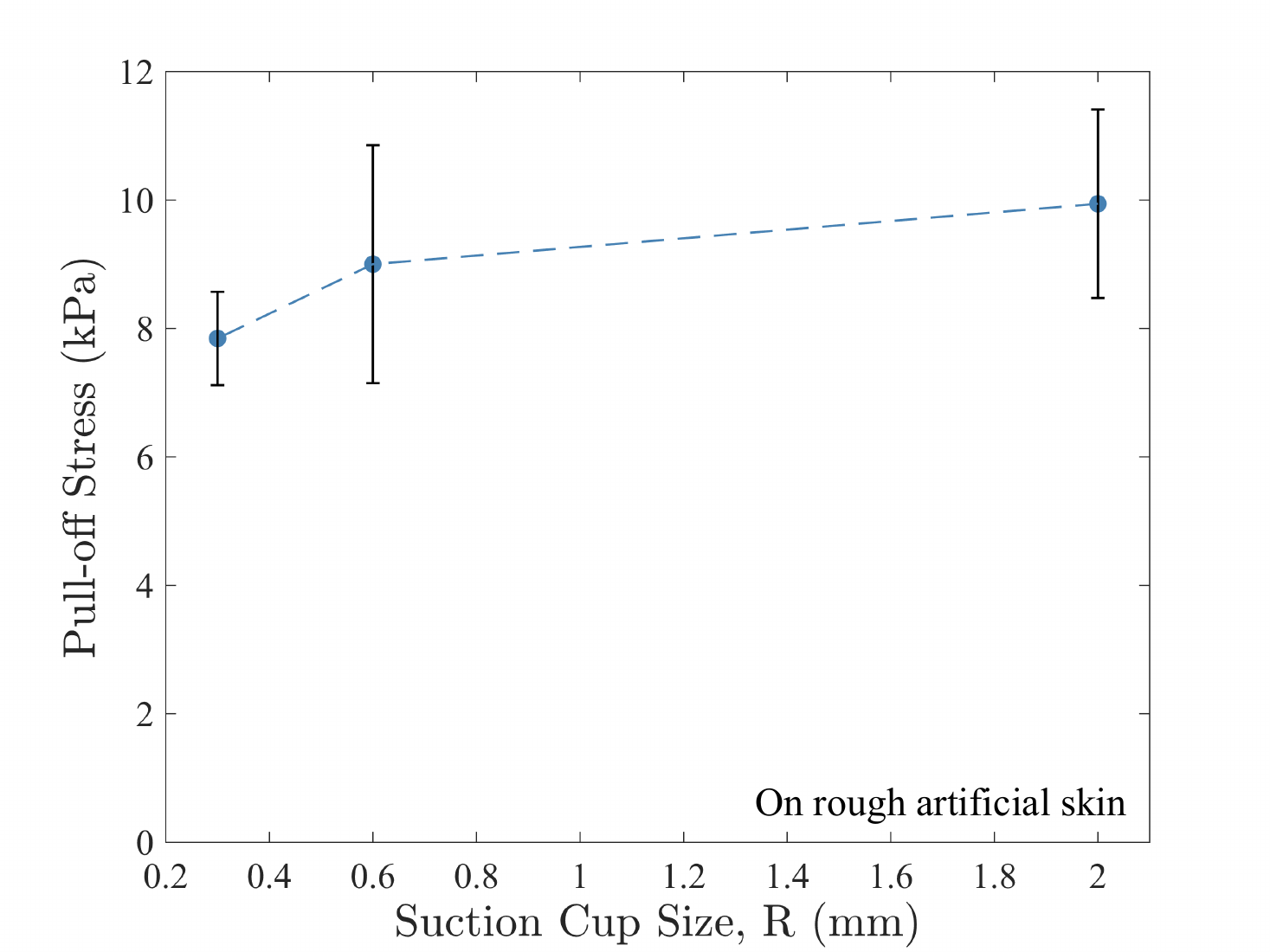}
    \caption{Suction cup arrays with varying cup radii (R) and fixed geometry ratios ($a = R$,$t = 0.3R$) tested on rough artificial skin.}
    \label{SIfig:normal histogram_aRt03R}
\end{figure}

\clearpage\section*{3. Suction cup modeling}
\begin{figure}[htb!]
    \centering
\includegraphics[width=0.95\textwidth]{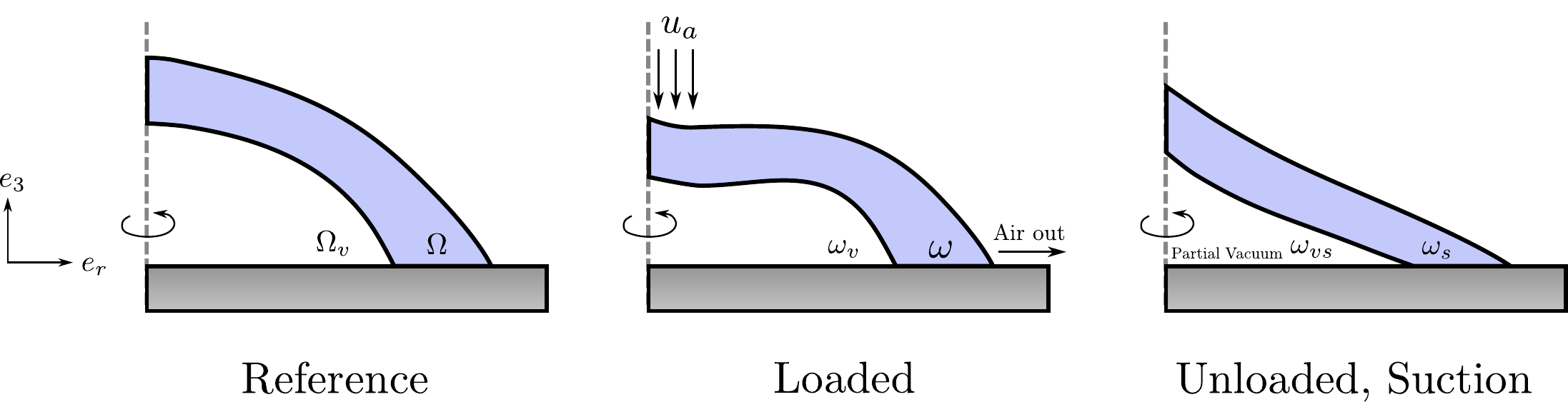}
    \caption{Diagram of the axisymmetric suction cut in the reference, loaded, and unloaded suction states.}
    \label{fig:fig1}
\end{figure}

We model the suction cup through a quasi-static, axisymmetric formulation. We consider two distinct loading states. In the first, the un-deformed cup structure is loaded while allowing air to escape the enclosed region. In the second phase, air may not enter or leave the cavity. Through the elastic recovery of the unloaded structure, a partial vacuum is formed in the enclosed region, as shown Figure~\ref{fig:fig1}. 
We first model the loading stage through large-deformation kinematics while allowing free variation of the enclosed volume. Then, we model the unloading vacuum stage by coupling the cavity volume with an internal pressure evolution through an equation of state.

\subsection{Loading Stage}
The formulation in the loading stage is a standard finite-elasticity problem. We consider an axisymmetric structure occupying $\Omega$ in the reference configuration that undergoes an applied displacement $u_a$ on its top boundary $\partial_u \Omega$. We consider finite-deformation kinematics with an internal energy density function $W^{cup}$. The free energy of the structure is the elastic energy of the cup,
\begin{equation}
    \mathcal{E}^L = \mathcal{E}^{el} \coloneq  2 \pi \int_\Omega W^{cup}(F) R \ d\Omega
\end{equation}
where $F(\nabla u, u_r)$ is the deformation gradient tensor, dependent on the displacement gradient $\nabla u$ and the radial displacement $u_r$ in the axisymmetric setting. For now, we leave the form of the energy function $W^{cup}$ general, and choose a particular constitutive law in the following sections. Taking variations gives the weak-form of equilibrium
\begin{equation}
    0 = 2 \pi \int_\Omega R \pdv{W^{cup}}{F} \cdot \left( \pdv{F}{\nabla u} \cdot \nabla \delta u + \pdv{F}{u_r} \delta u_r  \right) \ d\Omega  \qquad \text{ for all } \delta u \in \mathcal{U}_0,
\end{equation}
where $\mathcal{U}_0$ is the space of kinematically admissible displacement variations
\begin{equation}
    \mathcal{U}_0 := \left \{ u \in H^{1}(\Omega), \ u = 0 \text{ on } \partial_u \Omega \right \}, 
\end{equation}
where $H^1(\Omega)$ is the standard vector-valued Hilbert space.

\subsection{Suction Stage}
\begin{figure}[htb!]
    \centering
\includegraphics[width=0.75\textwidth]{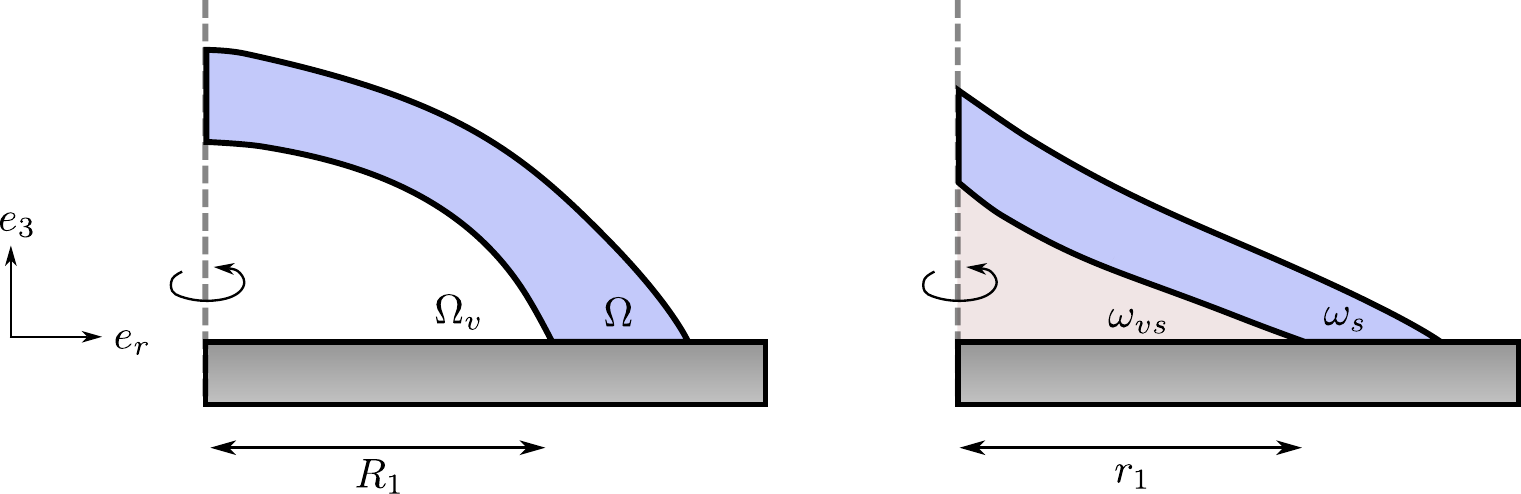}
    \caption{Diagram showing the reference and unloaded suction regions.}
    \label{fig:fig2}
\end{figure}

In the suction stage, careful consideration must be made when handling the pressure and volume in the enclosed region. We consider the isothermal work done by the structure onto the gas in the enclosed region as
\begin{equation}
    \mathcal{E}^{gas} \coloneq -\int_{V_0}^V (p - p_0) \ dV,
\end{equation}
where $p$ is the pressure in the enclosed region, $p_0$ is the ambient pressure, $V_0$ is the enclosed volume at the initiation of the suction stage, and $V$ is the current volume of the enclosed region. Assuming an ideal gas, 
\begin{equation}
    p V = p_0 V_0.
\end{equation}
Then,
\begin{equation}
    \mathcal{E}^{gas} = - \int_{V_0}^V \left( \frac{p_0 V_0}{V} - p_0\right) \ dV = - p_0 V_0 \log{\frac{V}{V_0}} + p_0 (V - V_0).
\end{equation}
The total energy of the system in the suction phase is the sum of the elastic energy of the cup and the work on the gas,
\begin{equation} \label{eq:energy_pv}
    \mathcal{E}^{S} = \mathcal{E}^{el} + \mathcal{E}^{gas} = 2 \pi \int_{\Omega} W^{cup}(F) R \ d\Omega - p_0 V_0 \log{\frac{V}{V_0}} + p_0 (V - V_0). 
\end{equation}
where $V$ is the deformed volume of the enclosed region $\omega_{vs}$,
\begin{equation}
    V := \int_{\omega_{vs} } 2 \pi r \ d\omega 
\end{equation}
in the cylindrical coordinate system assuming axisymmetry. We pull this back to the reference configuration as
\begin{equation} \label{eq:enclosed_vol}
    V = \int_{\Omega_{v} } \det(F_{2D}) 2 \pi (R + u_r) \ d\Omega,
\end{equation}
where $F_{2D}$ is the 2D deformation gradient for the mapping from $\Omega_v$ to $\omega_{vs}$. We may insert this relation into~\eqref{eq:energy_pv} and take variations to obtain the weak form of equilibrium. However, this system is highly non-linear, with the equation of state being directly coupled to the expression for the enclosed volume. This system is difficult to solve, and may not converge with a standard Newton-Raphson scheme. To mitigate this, we reintroduce the pressure $p$, which acts as a Lagrange multiplier constraining an auxiliary variable $\bar{V}$ to the enclosed volume. This changes the system energy to
\begin{equation}\mathcal{E}^{S} = 2 \pi \int_{\Omega} W^{cup}(F) R \ d\Omega - p_0 V_0 \log{\frac{\bar{V}}{V_0}} + p_0 (V - V_0) - p (V - \bar{V}). 
\end{equation}
Then, rearranging terms gives
\begin{equation}
        \mathcal{E}^{S} = 2 \pi \int_{\Omega} W^{cup}(F) R \ d\Omega - p_0 V_0 \log{\frac{\bar{V}}{V_0}} - (p - p_0) (V - \bar{V}) - p_0 (V_0 - \bar{V}). 
\end{equation}
Substituting~\eqref{eq:enclosed_vol} in for $V$ and taking variations with $u$, $p$, and $\bar{V}$ gives the equilibrium relations
\begin{equation} \label{eq:equil_sucky}
\begin{aligned}
        0 &= 2 \pi \int_\Omega R \pdv{W^{cup}}{F} \cdot \left( \pdv{F}{\nabla u} \cdot \nabla \delta u + \pdv{F}{u_r} \delta u_r  \right)  \ d\Omega  \\
        & \quad -  (p - p_0)  \int_{\Omega_v}  \det(F_{2D})  2 \pi  \left( \delta u_r + (R + u_r) F^{-T}_{2D} \cdot \nabla \delta u \right) \ d\Omega_v , \qquad \text{ for all } \delta u \in \mathcal{U}_0, \\
        0 &= \bar{V} - \int_{\Omega_{v} }  \det(F_{2D}) 2 \pi (R + u_r) \ d\Omega \\
        0 &=  p - \frac{p_0 V_0}{\bar{V}}. 
\end{aligned}
\end{equation}
The first equation is the balance of linear momentum, the second relation constrains the auxilliary volume $\bar{V}$ to the enclosed volume, and the last equation is the ideal gas law equation-of-state. Thus, the above is a non-linear set of equations for the displacement field, the pressure, and the volume in the enclosed region.
\subsection{Pull-off Condition}

In the suction stage, the top portion is incrementally pulled upwards until the pull-off condition is met. This occurs when the stretch in the cup overcomes the vacuum of the enclosed gas. In this case, it is when the net contact force between the cup and the substrate becomes non-compressive, that is, when tension must be applied to the bottom of the cup to maintain contact. We define this as when the net reaction force on the bottom of the cup is zero,
\begin{equation}
    0 = F_R = 2 \pi \int_{\partial_b \Omega} e_3 \cdot  \pdv{W^{cup}}{F} e_3  R \ dR,
\end{equation}
where $\partial_b \Omega$ is the boundary on the bottom of the cup. 

\subsection{Soft Substrates and Footing Layers}
For both the soft substrate and footing layer, we consider additional layers of soft, hyper-elastic materials. That is, we consider the elastic energy as
\begin{equation}
    \mathcal{E}^{el} := 2 \pi \left[ \int_\Omega  W^{cup}(F) R \ d\Omega + \int_{\Omega_f}  W^{foot}(F) R \ d\Omega_f  + \int_{\Omega_{sub}}  W^{sub}(F) R \ d\Omega_{sub} \right],
\end{equation}
where $W^{foot}$ and $W^{sub}$ are the energy density functions of the footing layer and the substrate. Then, the equilibrium condition is derived in the same manner as detailed in the previous section.

\subsection{Numerics and Solution Strategy}
The energy in~\eqref{eq:energy_pv} does not have any elastic contributions in the enclosed region $\Omega_v$. This leads to an ill-posed problem, as an infinite number of deformations in this region may give the same enclosed volume $V$. A standard method to overcome this is to introduce a very soft elastic solid in the enclosed region. We consider an additional elastic energy in the enclosed region 
\begin{equation}
    \mathcal{E}^{enc} = 2 \pi \int_{\Omega_v} W^{enc}(F) R \ d\Omega,
\end{equation}
with $W^{enc}$ being much softer than $W^{cup}$. For our computations, we consider a compressible Neo-Hookean constitutive law with a Poisson's ratio $\nu = 0.495$ for the cup and substrate. The solid in the vacuum regions is also considered near-incompressible with $\nu_v = 0.495$ to prevent excessive deformation. With $\mu$ being the shear modulus of the cup, we consider a shear modulus in the vacuum region of $\mu_v = 10^{-5} \mu$, which is found to have a negligible effect on the internal cup pressure. 

To solve the equilibrium relations~\eqref{eq:equil_sucky}, we consider a finite element formulation with standard $Q = 1$ quadrilateral elements. We mesh the entirety of the domain, including the cup, the vacuum region, and the substrate. We solve the system of equations with fully coupled Newton-Raphson iterations. That is, considering the displacement field, pressure, and volume as the unknowns to solve the system of equilibrium relations.

\clearpage\section*{4. Preparation and surface roughness of artificial skin}
\begin{figure}[htb!]
    \centering
\includegraphics[width=0.85\textwidth]{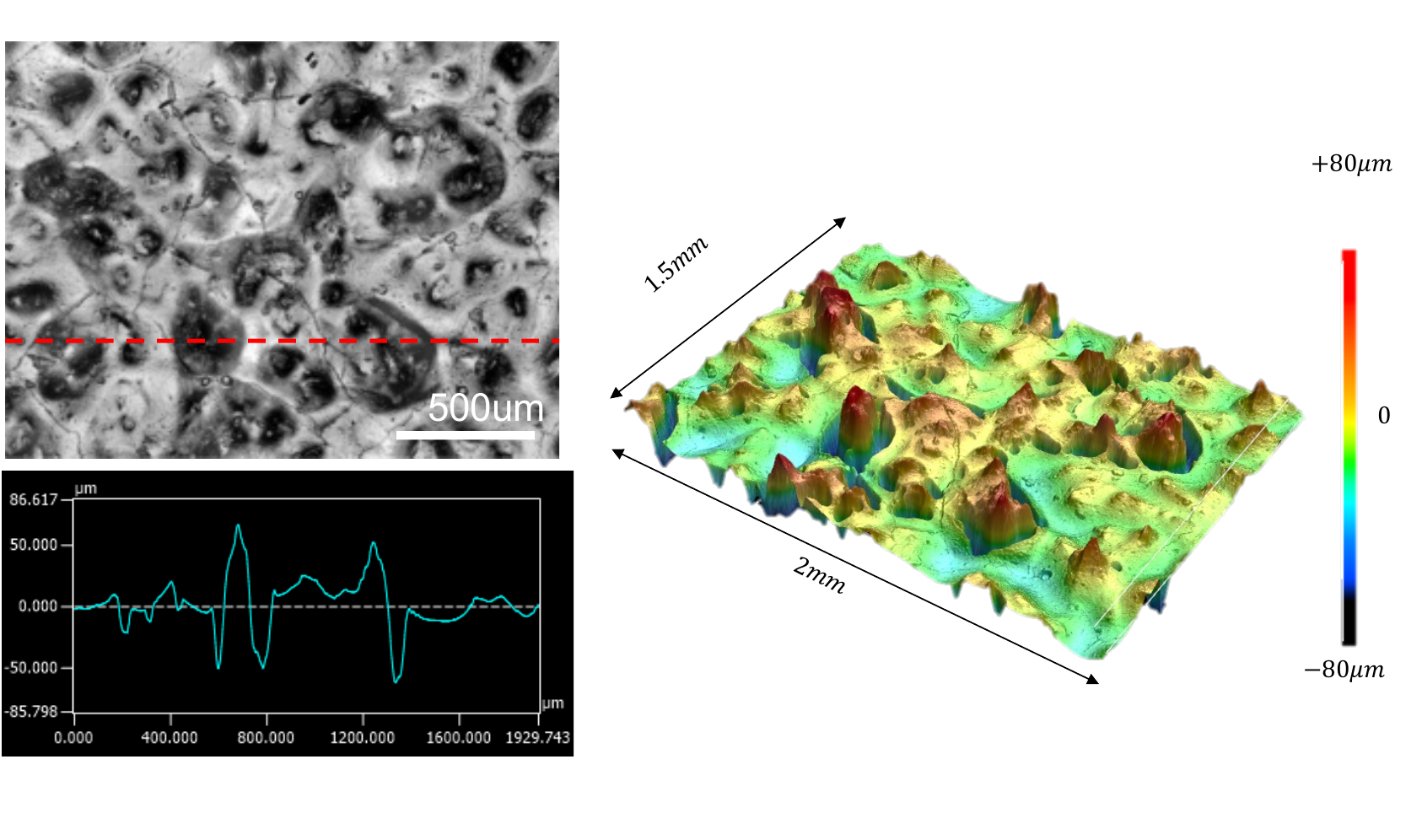}
    \caption{Confocal microscope scan of artificial skin (Vitro Skin) for roughness mapping}
    \label{SIfig:roughness mapping}
\end{figure}

\clearpage\section*{5. Gelsight measurement for substrate deformation under suction}
The setup consists of a displacement controlled linear actuator positioned on the left side of the system, responsible for providing precise transitional motion. Attached to the actuator, is a high precision Nano 17 force sensor which enables the real time monitoring of the compressive load applied during the testing. Connected to the force sensor is a flat surface compressor which interfaces with a suction cup sample. This suction cup is mounted directly onto the fine tactile sensing gel substrate part of the Gelsight to capture high resolution scans of the deformation.
\begin{figure}[htb!]
    \centering
\includegraphics[width=0.65\textwidth]{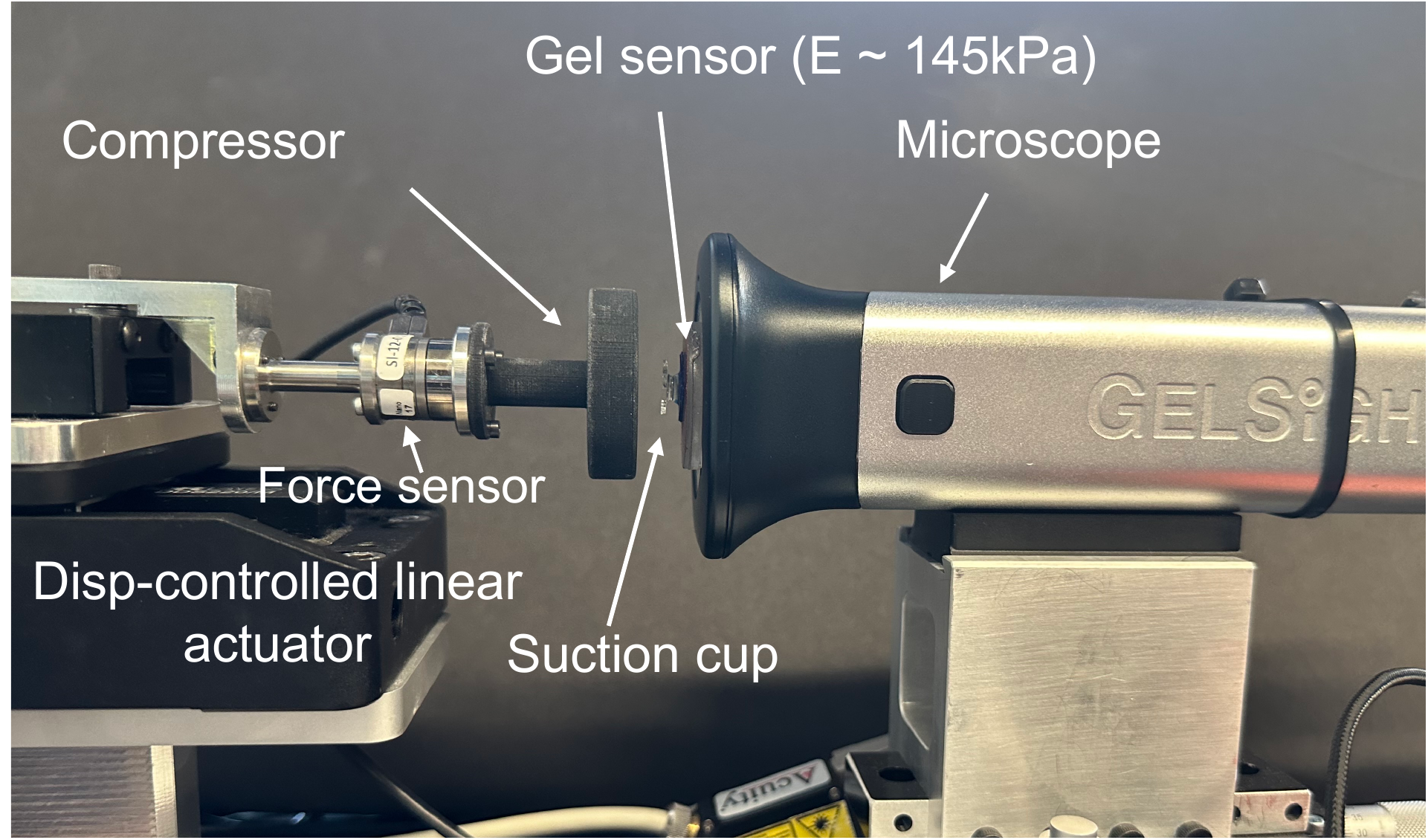}
    \caption{Gelsight measurement set-up}
    \label{SIfig:gelsight setup}
\end{figure}
During the operation, the linear actuator advances the compressor toward the suction cup mounted on the Gelsight sensor. As the compressor contacts the suction cup, the force sensor continuously measures the applied normal force. Once force reaches a predetermined threshold value, the actuator stops the motion and retracts rapidly to allow the cup deformation recovery and vacuum generation within the cup chamber. At this point, Gelsight captures a detailed scan of the substrate deformation, enabling high-fidelity analysis of the mechanical interactions between the substrate and the suction cup sample.

\begin{figure}[htb!]
    \centering
\includegraphics[width=0.85\textwidth]{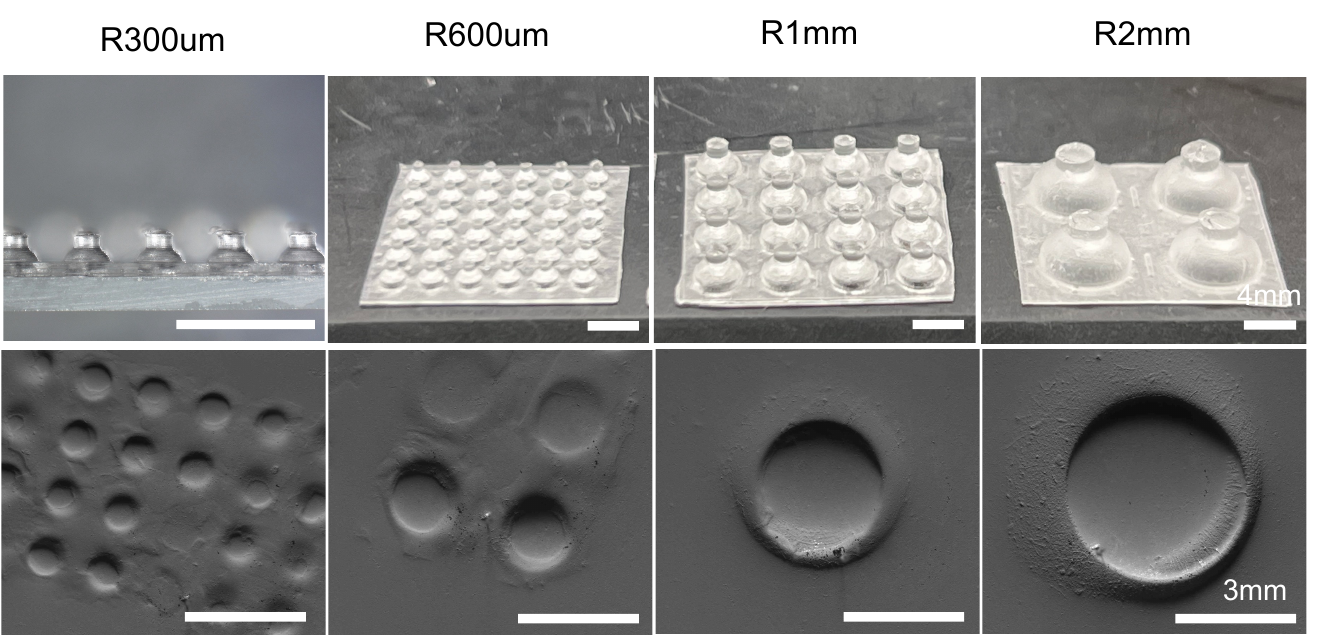}
    \caption{GelSight measurement of substrate deformation induced by vacuum cup arrays. Top images show optical views of suction cups with different radii ($R$). Scale bar 4 mm. Bottom images show the corresponding GelSight deformation profiles generated by each cup. Scale bar: 3 mm.}
    \label{SIfig:cup_array_gelsight}
\end{figure}

\begin{figure}[htb!]
    \centering
\includegraphics[width=0.95\textwidth]{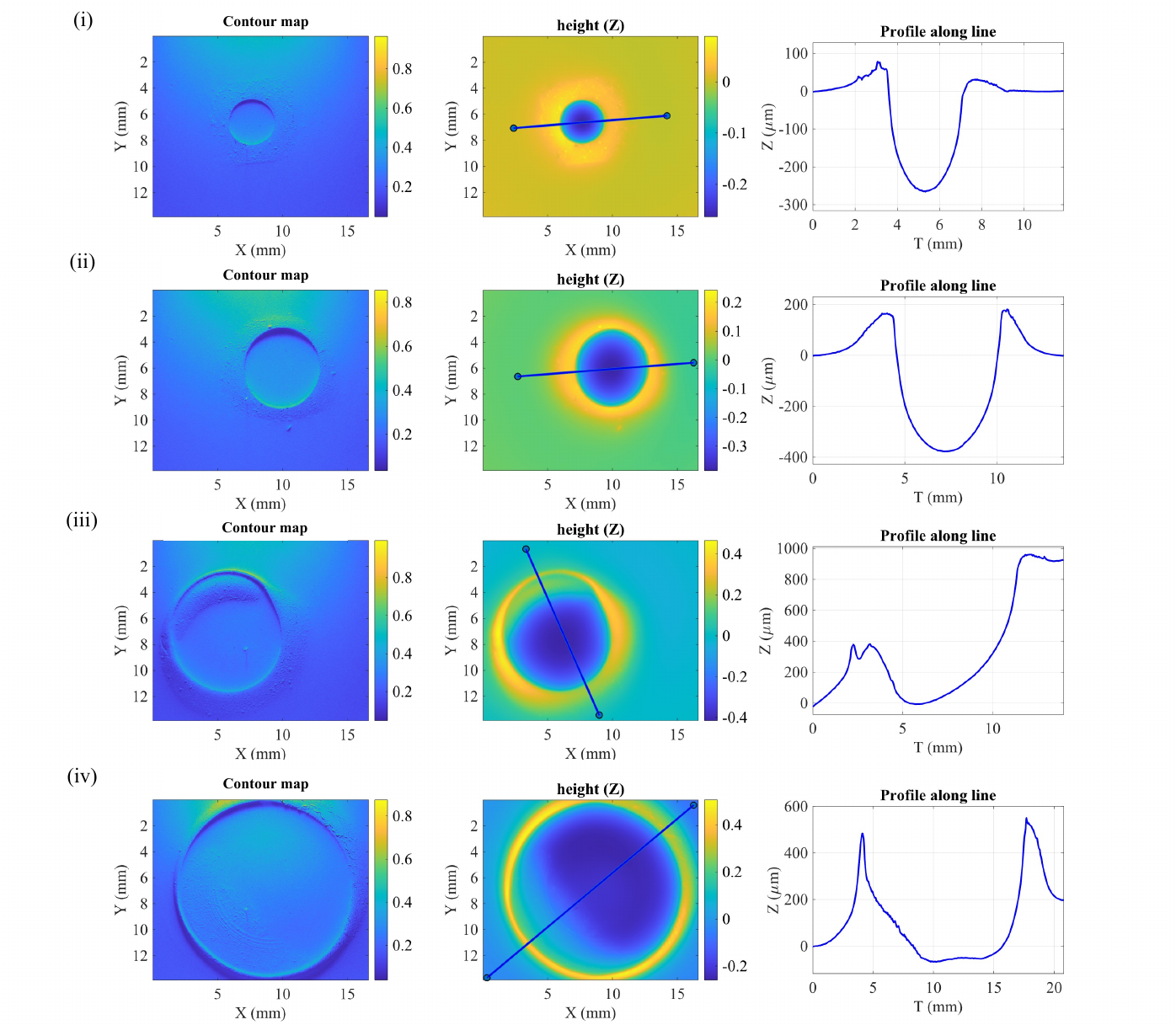}
    \caption{Effects of cup width on compliant surface deformation. Gelsight measurement and analysis for cups with fixed $R = 2mm$, t$ = 0.3R$ and varied $a$: (i) $a=0R$, (ii) $a=0.5R$, (iii) $a=1R$, (iv) $a=2R$.}
    \label{SIfig:gelsight data_a}
\end{figure}

\begin{figure}[htb!]
    \centering
\includegraphics[width=0.95\textwidth]{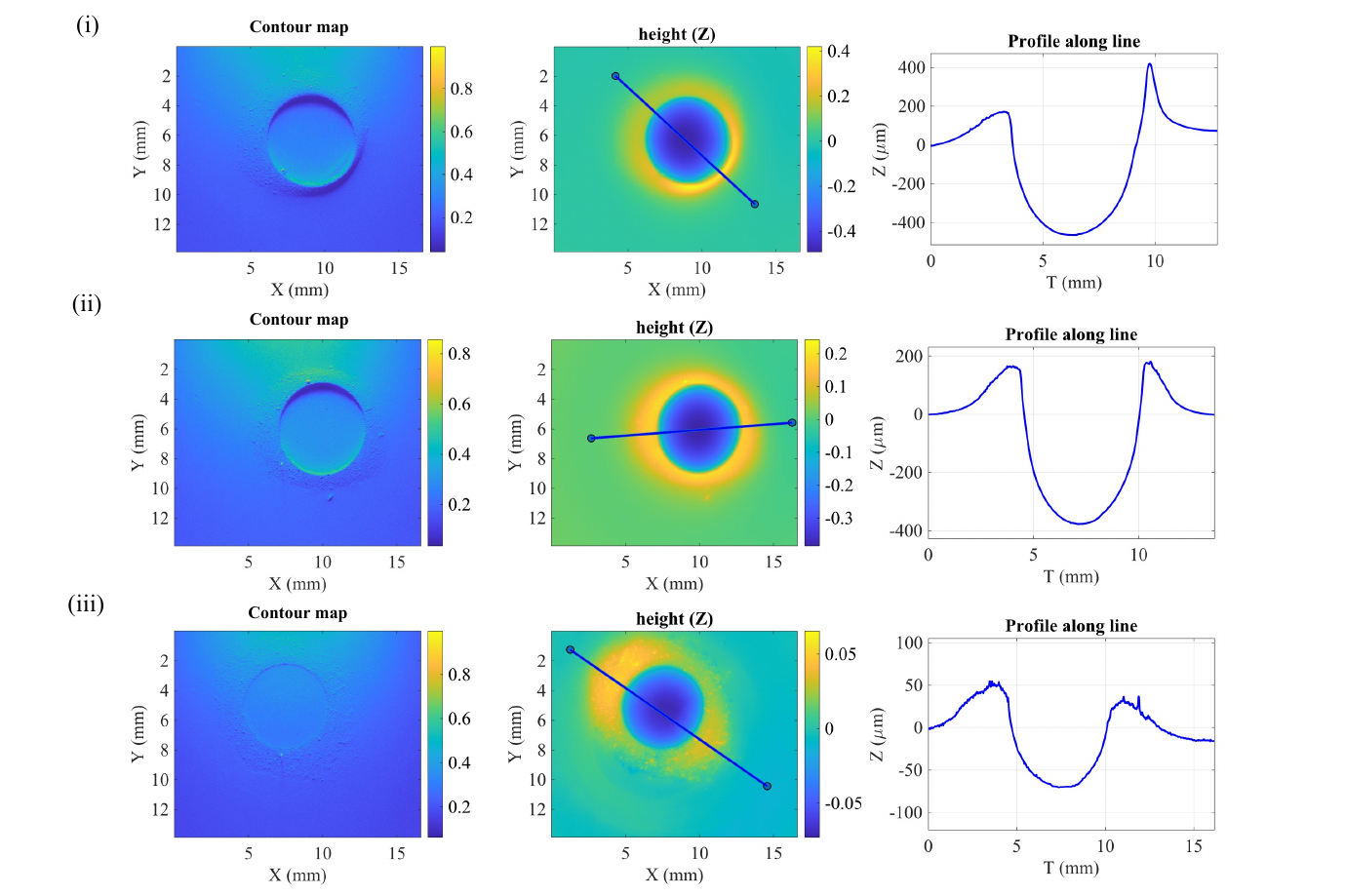}
    \caption{Effects of cup thickness on compliant surface deformation. Gelsight measurement and analysis for cups with fixed $R = 2mm$, $a = 0.5R$ and varied $t$: (i) $t=0.2R$, (ii) $t=0.3R$, (iii) $t=0.4R$.}
    \label{SIfig:gelsight data_t}
\end{figure}

\clearpage\section*{6. Work of adhesion measurement for interfacial materials}
\begin{figure}[htb!]
    \centering
\includegraphics[width=0.95\textwidth]{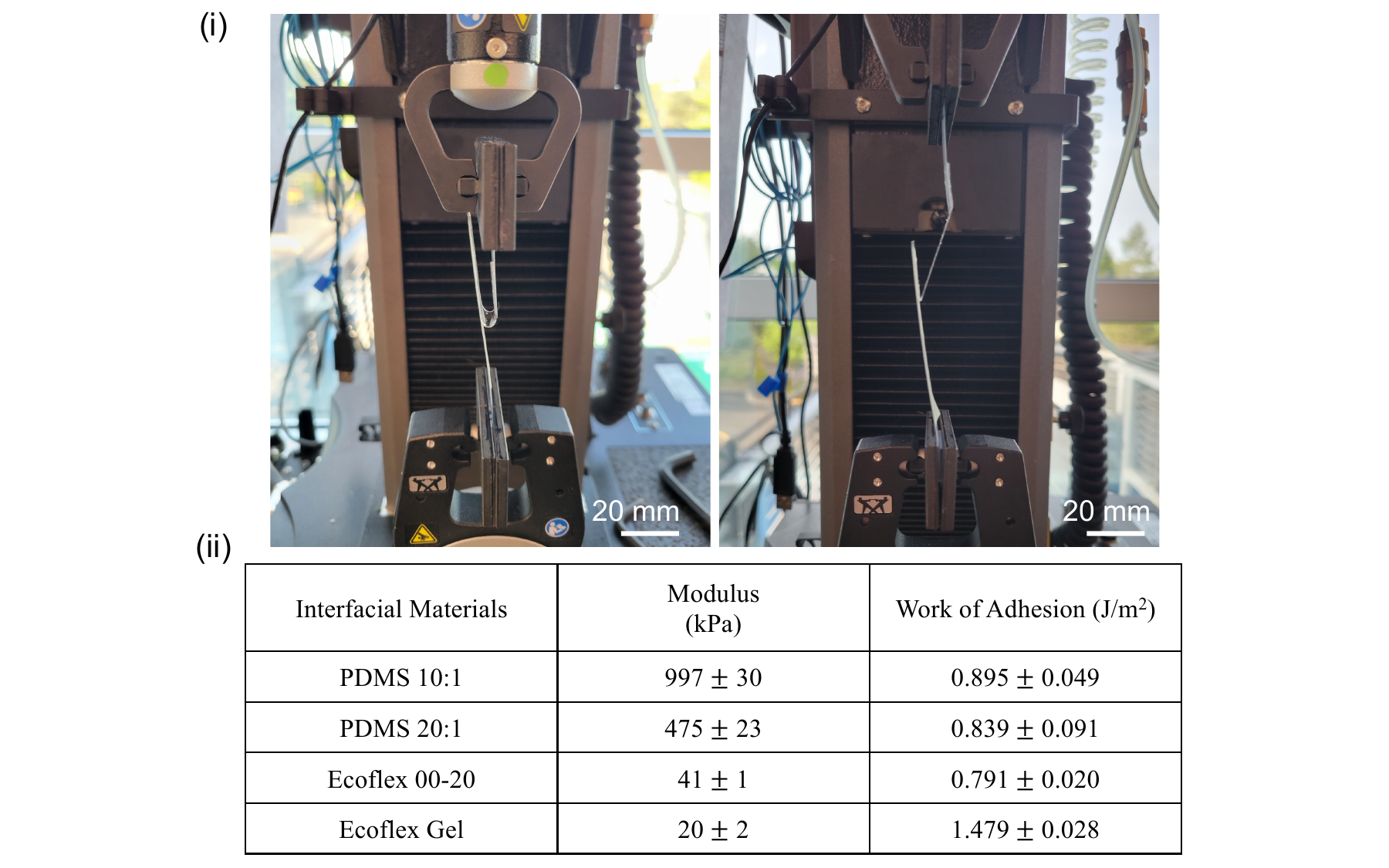}
    \caption{(i)Experimental set-up for the peel test to characterize the work of adhesion between interface materials and the artificial skin substrate. (ii) Summary of measured moduli and adhesion energies for each material. Values represent the mean and standard deviation from three independent samples per condition.}
    \label{SIfig:WOA}
\end{figure}

\clearpage\section*{7. Real contact measurement}
\begin{figure}[htb!]
    \centering
\includegraphics[width=0.65\textwidth]{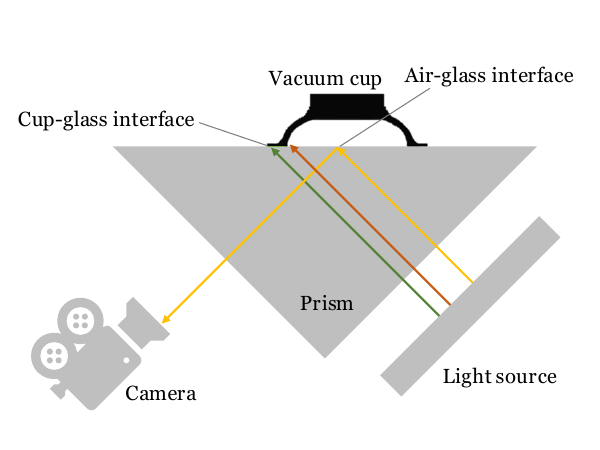}
    \caption{Real contact area measurement setup.}
    \label{SIfig:Real_CA_Cups}
\end{figure}
The setup for contact area measurement is based on total internal reflection of light when traveling from a glass with high refractive index ($n_i$ = 1.51509 for N-BK7 at 632.8nm) to air with low refractive index ($n_r$ = 1.00028 at 632.8nm) at an angle that is higher than the critical angle given by sin $\theta_{air} = n_r/n_i~ \rightarrow~\theta_{air}= 41.3^\circ$. As shown in the schematic of the measurement setup in Fig. \ref{SIfig:Real_CA_Cups}, the light from the white collimated light source hits the hypotenuse of the right-angle prism at 45$^\circ$, larger than the critical angle $\theta_{air}= 41.3^\circ$, and therefore reflects and arrives at the camera. However, when a surface such as vacuum cup contacts the glass surface, at the interface between glass and vacuum cup the light does not reflect and gets absorbed by the vacuum cup, and therefore the camera sees these contact areas as darker regions, having the same color as the vacuum cup. The setup is made using a right-angle prism with N-BK7 glass (50 mm, Uncoated, N-BK7 Right Angle Prism, Stock \#32-535, Edmund Optics), collimated white backlight (2"$\times$2" White Metaphase Technologies Collimated LED Backlight, \#37-082, Edmund Optics), off-the-shelf camera (MOKOSE 4K@30fps USB Camera with 6-12 mm Varifocal Manual Lens Webcam UVC), and a custom 3D-printed fixture.

\clearpage\section*{8. ECG/EMG experiment and data processing}
\begin{figure}[htb!]
    \centering
\includegraphics[width=0.5\textwidth]{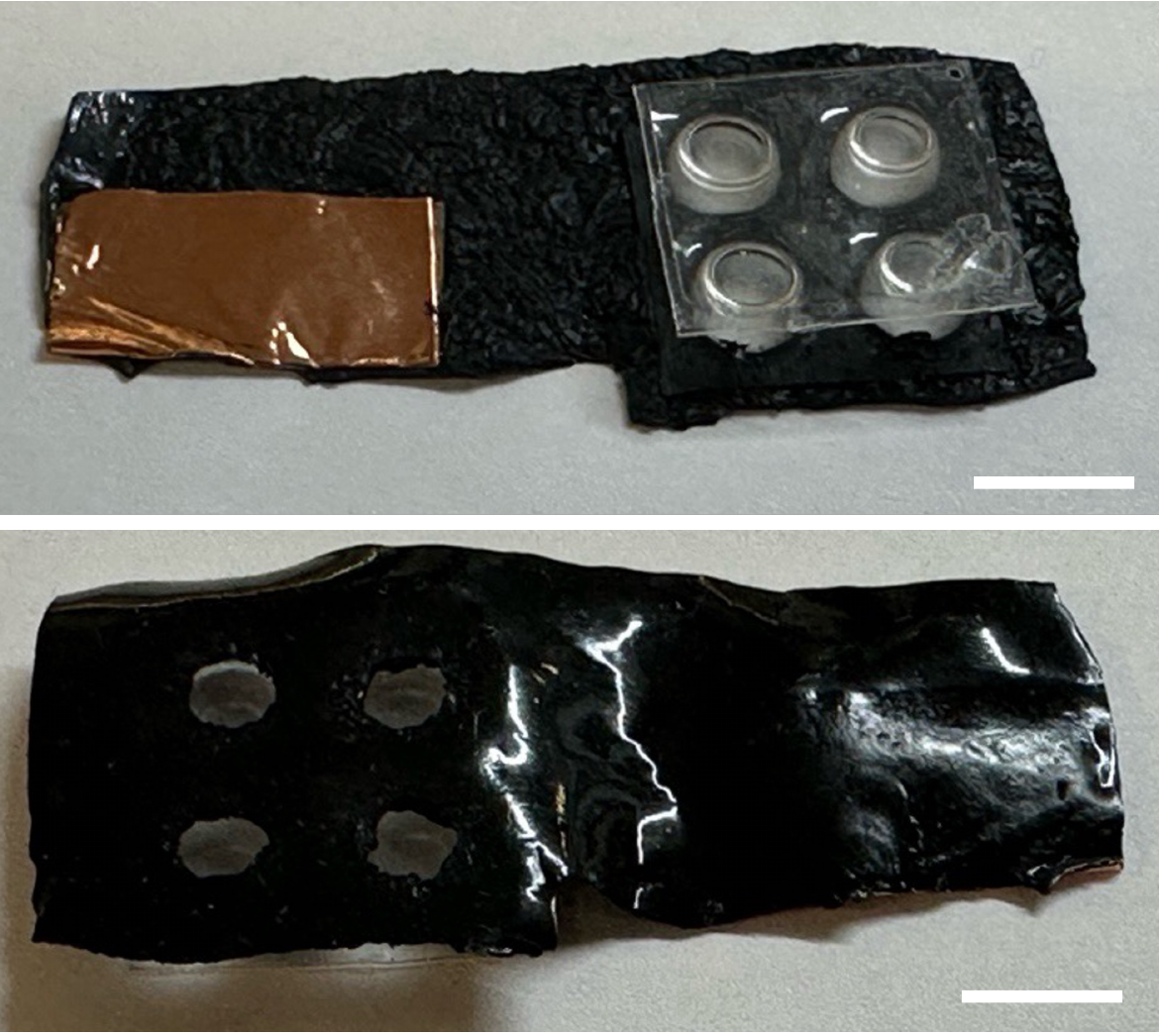}
    \caption{2$\times$2 suction cup array with $R=2mm$, $a=0.5R$. and $t=0.3R$, coated with a conductive footing layer for ECG/EMG measurement. Scale bar 10mm}
    \label{SIfig:ECG_EMG_cups}
\end{figure}

The conductive footing layer is intentionally made significantly larger than the size of the vacuum cup array for electrical connection as shown in Fig. \ref{SIfig:ECG_EMG_cups}. A piece of copper tape is taped on a portion of the dangling footing layer to make it locally stiffer and increase the area of electrical connection. Then the connectors from the ECG/EMG wires are clamped onto the copper tape as shown in Fig. 4 of the main paper. ECG and EMG biosignals are recorded from the Analog Front End (AFE) of a commercial wireless surface EMG system (Ultium EMG, Noraxon USA Inc., Scottsdale, AZ) with a built-in 24 bit ADC that has a sensitivity of 300 nV resolution. A sampling rate of 4000 Hz is programmed during the entire data collection.
For EMG measurement, an EMG sensor is attached to the body as a ground reference, a pair of soft electrodes with vacuum cups are attached to the forearm/wrist area to form a bipolar sensing. Then the user performs various muscle activities (such as maximum voluntary contraction, index-thumb finger pinch and release) while EMG signals are recorded and post-processed. The EMG envelope is post-processed via RMS with a 500 ms moving window and a bandpass filter of 10 to 1000 Hz. ECG measurement setup follows the dual-wrist ECG configuration, where the two sensing electrodes are placed on separate hands on each side of the heart so that cardiac action potential can be measured. Specifically, the ground reference and one sensing electrode are placed on the left wrist and the other sensing electrode is placed on the right wrist. The ECG raw signal is post-processed with a bandpass filter of 0.05 to 300 Hz.

\clearpage\section*{9. VR system set-up}
\begin{figure}[htb!]
    \centering
\includegraphics[width=0.95\textwidth]{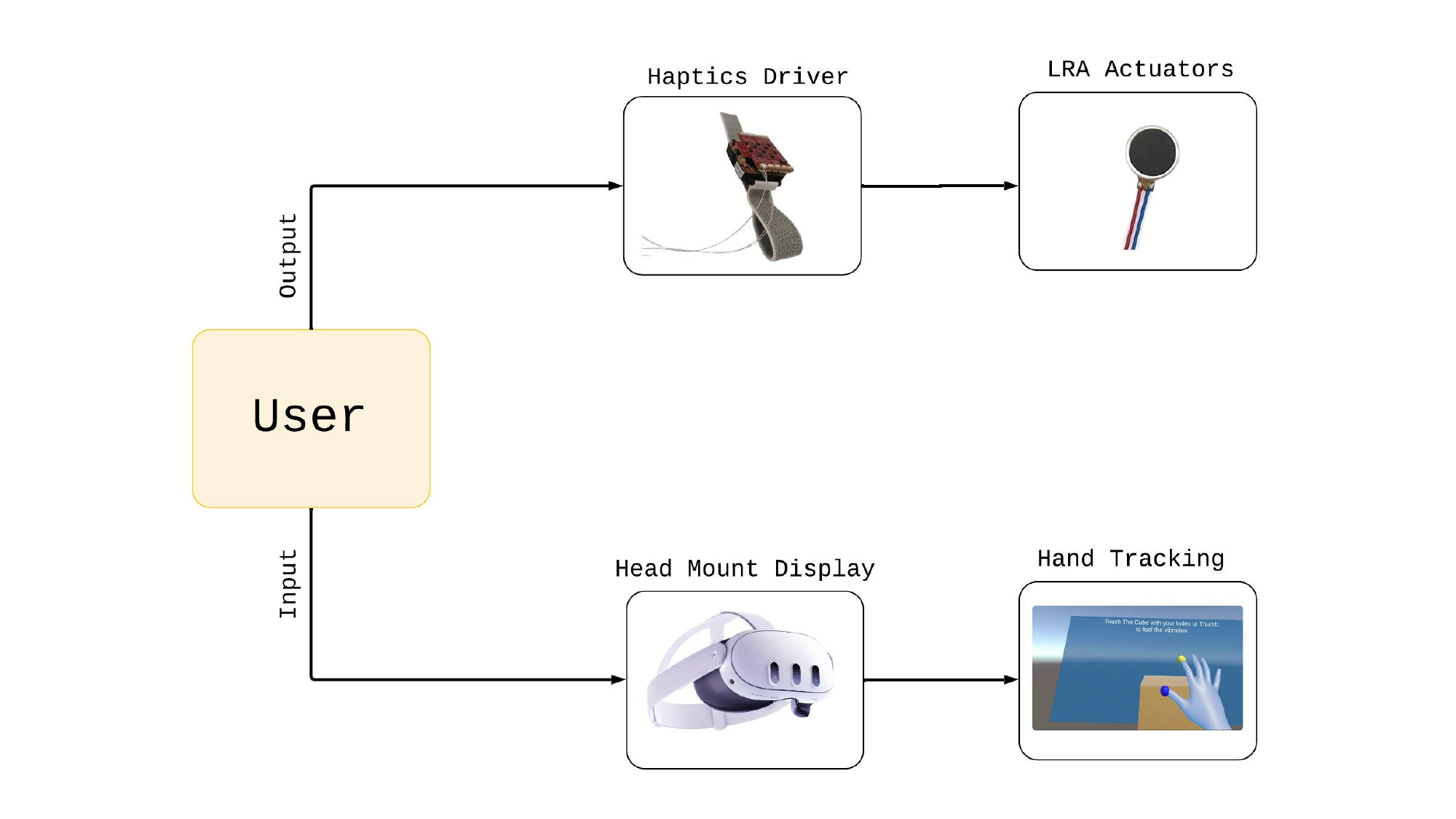}
    \caption{VR setup for tactile feedback. User wears head-mounted display with hand tracking. Touching the floating cube triggers haptic feedback via LRAs.}
    \label{SIfig:VR_setup}
\end{figure}
As an example of a use case, we develop a simple virtual reality environment that enables users to interact with virtual objects and receive tactile haptic feedback produced by actuators that are mounted on the user’s hand via our suction cups. The environment includes a floating cube that can be interacted with using the thumb and index fingers. Users wear a wristband that contains the haptic driver connected to two Linear Resonant Actuators (LRAs), which are attached to the index finger and thumb via suction cups. When the user touches the floating cube with their thumb or index finger in the virtual environment, the corresponding actuator is activated through the driver, providing a sense of tactile response.

\clearpage
%%% Add this line AFTER all your figures and tables
\FloatBarrier

\section*{10. Movies}
\begin{enumerate}
    \item Double-sided vacuum cup arrays for weight lifting by optimizing cup geometries for hard and soft substrates. \\ 
    \textbf{Movie S1.mp4}
    \item Vacuum cup integration with an inertial measurement unit (IMU) for motion sensing on human fingernails. \\
    \textbf{Movie S2.mp4}
    \item Vacuum cup-mounted linear resonant actuators (LRA) delivering localized haptic feedback during virtual reality interaction. \\
    \textbf{Movie S3.mp4}
    \item Enhanced contact of a wearable wristband using vacuum cups. \\
    \textbf{Movie S4.mp4}
    \item Electromyography (EMG) signal acquisition using suction cups with conductive footing layers. \\
    \textbf{Movie S5.mp4}
\end{enumerate}

% \dataset{dataset_one.txt}{Type or paste legend here.}

%\bibliography{pnas-sample}